\documentclass[fleqn]{lmcs}
\pdfoutput=1

\usepackage{lastpage}
\lmcsdoi{17}{1}{14}
\lmcsheading{}{\pageref{LastPage}}{}{}%
{Jan.~23,~2020}{Feb.~10,~2021}{}


\usepackage[utf8]{inputenc}
\usepackage{amsmath,amssymb,amsthm,stmaryrd,centernot,semantic,xspace,mathtools}
\usepackage{procalg}
\usepackage[normalem]{ulem}
\usepackage{relsize}
\usepackage{pgf}
\usepackage{tikz}
\usetikzlibrary{arrows,automata}
\usetikzlibrary{positioning}
\tikzset{
    state/.style={
           rectangle,
           rounded corners,
           draw=black, very thick,
           minimum height=2em,
           inner sep=2pt,
           text centered,
           minimum width=.9cm
           },
}

\usepackage{color}
\usepackage{ulem}

\newcommand{\T}{\ensuremath{\mathcal{T}} }
\newcommand{\A}{\ensuremath{\mathcal{A}}}
\newcommand{\Atau}{\ensuremath{\mathcal{A}_{\tau}}}
\newcommand{\Api}{\mathcal{A}_{\pi}}
\newcommand{\Sta}{\ensuremath{S}}
\newcommand{\Ins}{\ensuremath{\mathalpha{\uparrow}}}

\newcommand{\Atoms}{\ensuremath{\mathbb{A}}}
\newcommand{\R}{\mathrel{\mathcal{R}}}
\newcommand{\M}{\mathcal{M}}
\newcommand{\D}{\mathcal{D}}
\newcommand{\N}{\mathcal{N}}

\newcommand{\Dbox}{\mathcal{D}_{\Box}}

\newcommand{\tphdL}[1]{{#1 \!}^{\scriptscriptstyle <}}
\newcommand{\tphdR}[1]{\prescript{\scriptscriptstyle >}{}{\! #1}}
\newcommand{\Conf}[1][]{\mathalpha{\mathit{Conf}}_{#1}}

\renewcommand{\step}[2][]{\ensuremath{\mathbin{\arrow{#2}_{#1}}}}
  \newdimen\boxwdplusemdimen
  \def\arrow#1{{
     \boxwdplusemdimen=1em%
     \setbox0=\hbox{$\scriptstyle#1$}%
     \advance\boxwdplusemdimen by \wd0\relax%
     \ifdim\boxwdplusemdimen<16.11119pt%
       \boxwdplusemdimen=16.11119pt%
     \fi%
     \buildrel{#1}\over%
       {\setbox1=\hbox to \boxwdplusemdimen{\rightarrowfill}%
     \ht1=0.3em\relax\box1}%
   }}
\makeatletter
   \def\twoheadrightarrowfill{$\m@th\smash-\mkern-7mu%
     \cleaders\hbox{$\mkern-2mu\smash-\mkern-2mu$}\hfill
     \mkern-7mu\mathord\twoheadrightarrow$}
\makeatother
   \def\darrow#1{{
     \boxwdplusemdimen=1em%
     \setbox0=\hbox{$\scriptstyle#1$}%
     \advance\boxwdplusemdimen by \wd0\relax%
     \ifdim\boxwdplusemdimen<16.11119pt%
       \boxwdplusemdimen=16.11119pt%
     \fi%
     \buildrel{#1}\over%
       {\setbox1=\hbox to \boxwdplusemdimen{\twoheadrightarrowfill}%
     \ht1=0.3em\relax\box1}%
   }}

\newcommand{\bbisimd}{\ensuremath{\mathrel{\bbisim^{\Delta}}}}

\newcommand{\nil}{\ensuremath{\mathalpha{\mathbf{0}}}}

\newcommand{\outcap}[2]{\ensuremath{\mathalpha{\overline{#1}\,{#2}}}}
\newcommand{\incap}[2]{\ensuremath{\mathalpha{{#1}(#2)}}}
\newcommand{\taucap}{\ensuremath{\mathalpha{\tau}}}
\newcommand{\pref}[1]{\ensuremath{\mathalpha{#1}.}}
\renewcommand{\parc}{\ensuremath{\mathbin{\mid}}}
\newcommand{\restr}[2]{\ensuremath{\mathop{({\nu}#1}){#2}}}

\newcommand{\repl}[1]{\ensuremath{\mathalpha{!}#1}}

\newcommand{\ddef}{\overset{\textrm{def}}{=}}

\newcommand{\fn}[1]{\ensuremath{\mathsf{fn}(#1)}}
\newcommand{\bn}[1]{\ensuremath{\mathsf{bn}(#1)}}

\newcommand{\piact}[1][\alpha]{\ensuremath{\mathalpha{#1}}}
\newcommand{\inact}[2]{\ensuremath{\mathalpha{{#1}\,#2}}}
\newcommand{\outact}[2]{\ensuremath{\mathalpha{\overline{#1}\,{#2}}}}
\newcommand{\boutact}[2]{\ensuremath{\mathalpha{\overline{#1}\,(#2)}}}
\newcommand{\tauact}{\ensuremath{\mathalpha{\tau}}}

\newcommand{\delete}[1]{}

\newcommand{\defeqn}{\mathrel{\stackrel{\text{def}}{=}}}
\newcommand{\scong}{\mathrel{\equiv}}
\newcommand{\dcomp}{\mathrel{\rightsquigarrow}}

\newcommand{\fullversion}[1]{#1}
\newcommand{\iceonly}[1]{}

\newcommand{\longproofs}[1]{}

\begin{document}

\title{The \texorpdfstring{$\pi$}{pi}-Calculus is Behaviourally Complete and Orbit-Finitely Executable}

\author[B.~Luttik]{Bas Luttik}	
\address{Department of Mathematics and Computer Science, Eindhoven Unversity of Technology, The Netherlands}	
\email{s.p.luttik@tue.nl}  

\author[F.~Yang]{Fei Yang}	
\address{Zhejiang Lab, Hangzhou, China}	
\email{yangf@zhejianglab.com}  





\begin{abstract}
Reactive Turing machines extend classical Turing machines with a facility to model observable interactive behaviour. We call a behaviour (finitely) executable if, and only if, it is equivalent to the behaviour of a (finite) reactive Turing machine. In this paper, we study the relationship between executable behaviour and behaviour that can be specified in the $\pi$-calculus. We establish that every finitely executable behaviour can be specified in the $\pi$-calculus up to divergence-preserving branching bisimilarity. The converse, however, is not true due to (intended) limitations of the model of reactive Turing machines. That is, the $\pi$-calculus allows the specification of behaviour that is not finitely executable up to divergence-preserving branching bisimilarity. We shall prove, however, that if the finiteness requirement on reactive Turing machines and the associated notion of executability is relaxed to orbit-finiteness, then the $\pi$-calculus is executable up to (divergence-insensitive) branching bisimilarity.
\end{abstract}

\maketitle

\section{Introduction}\label{sec:intro}

In 2006, Jos Baeten initiated a research programme to explore and strengthen the connections between the classical theory of automata and formal languages and concurrency theory, with the aim to establish a unified theory. Such a unified theory should, in particular, upgrade the classical theory of automata and formal languages with a treatment of interaction, and reconsider standard notions and results modulo some form of bisimilarity instead of language equivalence.

Thus, non-deterministic finite automata and pushdown automata, and their correspondences with regular and context-free grammars were explored in the context of branching bisimilarity \cite{BCLvT2009}. The expressiveness of regular expressions modulo bisimilarity was characterised \cite{BCG2007}, and the expressiveness of extensions of regular expressions with various forms of parallel composition were studied \cite{BLMvT16}. The latter culminated in a concurrency-theoretic variant of Kleene's theorem establishing the correspondence, modulo bisimilarity, between finite automata and regular expressions extended with ACP-style parallel composition and encapsulation. Also, the idea that a pushdown automaton is a finite automaton interacting with a stack was formalised \cite{BCvT2008}.

Finite automata and pushdown automata have a straightforward labelled transition system semantics, and are therefore directly amenable to investigation from a concurrency-theoretic perspective. Turing machines, on the other hand, are designed to compute mathematical functions and not to exhibit interactive behaviour: the input  for the computation is assumed to be present on the Turing machine tape at the start of the computation, output is what is left on the tape after the computation, and the computation process itself is deemed internal. To add interactivity, an extension of the Turing machine was proposed that associates an action with every computation step \cite{BLvT2013}. This so-called \emph{reactive} Turing machine does have a labelled transition system semantics and can be studied from a concurrency-theoretic perspective.

In the same way as Turing machines have been used to define which functions are effectively computable, reactive Turing machines can be used to define which processes can be executed by a computing system. A process, mathematically modelled as a labelled transition system, is \emph{executable} if it is behaviourally equivalent to the labelled transition system associated with a  reactive Turing machine. Thus, reactive Turing machines provide a way to characterise the \emph{absolute expressiveness} of a process calculus, by determining to what extent transition systems specified in the calculus are executable, and by determining to what extent executable transition systems can be specified in the calculus. If it is possible to specify every executable transition system in a process calculus, then we say that the process calculus is \emph{behaviourally complete}. Note that the behavioural equivalence is a parameter of this method: if a process calculus is not behaviourally complete up to some fine notion of behavioural equivalence (e.g., divergence-preserving branching bisimilarity), it may still be behaviourally complete up to some coarser notion of behavioural equivalence (e.g., the divergence-insensitive variant of branching bisimilarity).  The entire spectrum of behavioural equivalences (see van Glabbeek's seminal paper \cite{Glabbeek1993}) is at our disposal to draw precise conclusions.

Expressiveness questions have received ample attention in concurrency theory, especially in the context of the $\pi$-calculus (see, e.g., \cite{Gorla2010,Fu2010}), but mostly pertaining to so-called \emph{relative} expressiveness: Is there a transformation of expressions of one calculus to expressions of another calculus preserving certain behavioural properties? In this article, instead, we consider the \emph{absolute} expressiveness of the $\pi$-calculus using the tool of reactive Turing machine.

Our first contribution, which was already announced in \cite{LY15}, is a characterisation of the expressiveness of the $\pi$-calculus according to the method described above. We prove that the $\pi$-calculus is behaviourally complete up to divergence-preserving branching bisimilarity: every executable behaviour can be specified in the $\pi$-calculus up to divergence-preserving branching bisimilarity \cite{vGW1996,vGLT2009}, which is the finest behavioural equivalence discussed in \cite{Glabbeek1993}. Our proof explains how the behaviour of an arbitrary finite reactive Turing machine can be simulated by a $\pi$-calculus expression. The specification consists of a  component that specifies the behaviour of the tape memory, and a component that specifies the behaviour of the finite control of the reactive Turing machine under consideration. The specification of the behaviour of the tape memory is generic and elegantly uses the link mobility feature of the $\pi$-calculus.

We also prove that the converse is not true: it is possible to specify, in the $\pi$-calculus, transition systems that are not executable up to divergence-preserving branching bisimilarity. We shall analyse the discrepancy and identify two causes.

The first cause is that the $\pi$-calculus presupposes an infinite supply of names, which is technically essential both for the way input is modelled and for the way fresh name generation is implemented. The infinite supply of names in the $\pi$-calculus gives rise to an infinite alphabet of actions. The presupposed alphabet of actions of a finite reactive Turing machine is, however, purposely kept finite.  Allowing reactive Turing machines to have an infinite alphabet of actions only makes sense if the reactive Turing machine also has infinitely many states or infinitely many data symbols, and then, without any alternative restrictions we get an unrealistically expressive model of executability. In fact, every countable transition system is then executable up to divergence-preserving branching bisimilarity. We refer to \cite[Section 6.1]{Yan18} for an elaboration.

The second cause is that the transition system associated with a $\pi$-calculus term may have unbounded branching, even if it refers to only finitely many names. Transition systems with unbounded branching are not executable up to divergence-preserving branching bisimilarity, but unbounded branching behaviour can be simulated at the expense of sacrificing divergence preservation. In \cite{LY15}, we proved that $\pi$-calculus processes referring to only finitely many names are executable up to (the divergence insensitive variant of) branching bisimilarity.

Our second contribution is a refinement and generalisation of the aforementioned result presented in \cite{LY15} regarding the executability of the $\pi$-calculus. Building on the foundations laid by Gabbay and Pitts on \emph{nominal techniques} \cite{GP2002} and subsequent work by Boja\'nczyk et al.\ on Turing machines with atoms \cite{BKLT13}, we propose a notion of \emph{orbit-finite} reactive Turing machine and \emph{orbit-finite} executability. The components of an orbit-finite reactive Turing machine (i.e., its sets of states, transitions, data symbols, action alphabet) are allowed to be infinite, as long as they are finitely presentable. We argue that the $\pi$-calculus is orbit-finitely executable up to branching bisimilarity.

The paper is organised as follows. In Section~\ref{sec:prelims}, the basic definitions of labelled transition system, (divergence-preserving) branching bisimilarty, and reactive Turing machine are recapitulated, and we also recall the syntax and operational semantics of the $\pi$-calculus with replication. In Section~\ref{sec:simulation}, we prove that the $\pi$-calculus is behaviourally complete modulo divergence-preserving branching bisimilarity: a finite specification of reactive Turing machines in the $\pi$-calculus is proposed and verified. In Section~\ref{sec:ofexecpi}, we discuss the orbit-finite executability of transition systems associated with $\pi$-calculus processes. We define the notion of orbit-finite reactive Turing machine and show that every $\pi$-calculus term can be simulated up to the divergence insensitive version of branching bisimilarity.
The paper ends with some conclusions in Section~\ref{sec:conclusion}.

\section{Preliminaries} \label{sec:prelims}

\subsection{Labelled Transition System and Behavioural Equivalence}\label{subsec:behaviour}

The transition system is the central notion in the mathematical theory of discrete-event behaviour. It is parameterised by a set $\A$ of \emph{action symbols}, denoting the observable events of a system. We shall later impose extra restrictions on $\A$, e.g., requiring that it be finite or have a particular structure, but for now we let $\A$ be just an arbitrary abstract set. We extend $\A$ with a special symbol $\tau$, which intuitively denotes unobservable internal activity of the system. We shall abbreviate $\A \cup\{\tau\}$ by $\Atau$.

\begin{defi} \label{def:lts}
An \emph{$\Atau$-labelled transition system}  is a triple $(\Sta,\step{},\uparrow)$, where
\begin{enumerate}
    \item $\Sta$ is a set of \emph{states}; 
    \item ${\step{}}\subseteq\Sta\times\Atau\times\Sta$ is an $\Atau$-labelled \emph{transition relation}; and
    \item ${\uparrow}\in\Sta$ is the initial state.
    \end{enumerate}

\end{defi}

\noindent
Let $(\Sta,\step{},\uparrow)$ be an $\Atau$-labelled LTS. We shall usually write $s\step{a}t$ in lieu of $(s,a,t)\in{\step{}}$.
The set $\textit{Reach}(s)$ of states reachable from a state $s$ is defined by
\begin{multline*}
\textit{Reach}(s)= \\ \{s'\in \Sta\mid \exists n\geq 0,\, s_0,\ldots,s_n\in\Sta,\, a_1,\ldots,a_n\in\Atau.\, s=s_0\step{a_1}\cdots\step{a_n}s_n=s'\}
\enskip.
\end{multline*}

Transition systems can be used to give semantics to programming languages and process calculi. The standard method is to first associate with every program or process expression a transition system (its operational semantics), and then consider programs and process expressions modulo one of the many behavioural equivalences on transition systems that have been studied in the literature. In this paper, we shall use the notion of (divergence-preserving) branching bisimilarity \cite{vGW1996,Lut2020}, which is the finest behavioural equivalence discussed in~\cite{Glabbeek1993}, and also the coarsest behavioural equivalence that is a congruence for parallel composition and preserves CTL$^{*}_{-X}$ formulas~\cite{vGLT2009b}.

In the definition of (divergence-preserving) branching bisimilarity we need the following notation: let $\step{}$ be an $\Atau$-labelled transition relation on a set $\Sta$, and let $a\in\Atau$; we write $s\step{(a)}t$ for ``$s\step{a}t$ or $a=\tau$ and $s=t$''. Furthermore, we denote the transitive closure of $\step{\tau}$ by $\step{}^{+}$ and the reflexive-transitive closure of $\step{\tau}$ by $\step{}^{*}$.

\begin{defi} \label{def:bbisim}
  Let $T_1=(\Sta_1,\step{}_1,\uparrow_1)$ and $T_2=(\Sta_2,\step{}_2,\uparrow_2)$ be transition systems. A \emph{branching bisimulation} from $T_1$ to $T_2$ is a binary relation ${\R}\subseteq{\Sta_1\times\Sta_2}$ such that for all states $s_1$ and $s_2$, $s_1\R s_2$ implies
\begin{enumerate}
    \item if $s_1\step{a}_1s_1'$, then there exist $s_2',s_2''\in\Sta_2$, such that $s_2\step{}_2^{*}s_2''\step{(a)}s_2'$, $s_1\R s_2''$ and $s_1'\R s_2'$;
    \item if $s_2\step{a}_2s_2'$, then there exist $s_1',s_1''\in\Sta_1$, such that $s_1\step{}_1^{*}s_1''\step{(a)}s_1'$, $s_1''\R s_2$ and $s_1'\R s_2'$.
\end{enumerate}
The transition systems $T_1$ and $T_2$ are \emph{branching bisimilar} (notation: $T_1\bbisim T_2$) if there exists a branching bisimulation $\R$ from $T_1$ to $T_2$ s.t. $\uparrow_1\R\uparrow_2$.

A branching bisimulation $\R$ from $T_1$ to $T_2$ is \emph{divergence-preserving} if, for all states $s_1$ and $s_2$, $s_1\R s_2$ implies
\begin{enumerate}
\setcounter{enumi}{2}
    \item if there exists an infinite sequence $(s_{1,i})_{i\in\mathbb{N}}$ such that $s_1=s_{1,0},\,s_{1,i}\step{\tau}s_{1,i+1}$ and $s_{1,i}\R s_2$ for all $i\in\mathbb{N}$, then there exists a state $s_2'$ such that $s_2\step{}^{+}s_2'$ and $s_{1,i}\R s_2'$ for some $i\in\mathbb{N}$; and
    \item if there exists an infinite sequence $(s_{2,i})_{i\in\mathbb{N}}$ such that $s_2=s_{2,0},\,s_{2,i}\step{\tau}s_{2,i+1}$ and $s_1\R s_{2,i}$ for all $i\in\mathbb{N}$, then there exists a state $s_1'$ such that $s_1\step{}^{+}s_1'$ and $s_1'\R s_{2,i}$ for some $i\in\mathbb{N}$.
\end{enumerate}
The transition systems $T_1$ and $T_2$ are \emph{divergence-preserving branching bisimilar} (notation: $T_1\bbisim^{\Delta}T_2$) if there exists a divergence-preserving branching bisimulation $\R$ from $T_1$ to $T_2$ such that $\uparrow_1\R\uparrow_2$.
\end{defi}

\noindent
For two LTSs $T_1=(\Sta_1,\step{}_1,\uparrow_1)$ and $T_2=(\Sta_2,\step{}_2,\uparrow_2)$, $s_1\in\Sta_1$ and $s_2\in\Sta_2$, we write $s_1\bbisim s_2$ ($s_1\bbisimd s_2$) if there is a (divergence-preserving) branching bisimilarity from $T_1$ to $T_2$ relating $s_1$ and $s_2$. Thus, $\bbisim$ is a relation from the states of $T_1$ to the states of $T_2$, and it can be shown that it satisfies the conditions of Definition~\ref{def:bbisim}.
We can also write $s_1\bbisim s_2$ ($s_1\bbisimd s_2$) if $s_1$ and $s_2$ are states in a single LTS $T$ and related by a (divergence-preserving) branching bisimulation from $T$ to itself.

The relations $\bbisim$ and $\bbisimd$ are equivalence relations, both as relations on a single transition system, and as relations on a set of transition systems \cite{Basten1996,vGLT2009}.

Next we define the notion of bisimulation up to $\bbisim$. Note that we adapt a non-symmetric bisimulation up to relation, which is a useful tool to establish branching bisimilarity later.

\begin{defi}\label{def:up-to}
Let $T_1=(\Sta_1,\step{}_1,\uparrow_1)$ and $T_2=(\Sta_2,\step{}_2,\uparrow_2)$ be two transition systems. A relation ${\R}\subseteq{\Sta_1\times\Sta_2}$ is a bisimulation up to $\bbisim$ if, whenever $s_1\R s_2$, then for all $a\in \Atau$:
\begin{enumerate}
    \item if $s_1\step{}^{*}s_1''\step{a}s_1'$, with $s_1\bbisim s_1''$ and ${a\neq\tau}\vee{s_1''\not\bbisim s_1'}$, then there exists $s_2'$ such that $s_2\step{a}s_2'$, $s_1''\mathrel{\bbisim\mathrel{\circ}\mathrel{\R}}s_2$ and $s_1'\mathrel{\bbisim \mathrel{\circ} \mathrel{\R}} s_2'$; and
    \item if $s_2\step{a}s_2'$, then there exist $s_1',s_1''$ such that $s_1\step{}^{*}s_1''\step{a}s_1'$, $s_1''\bbisim s_1$ and $s_1'\mathrel{\bbisim \mathrel{\circ} \mathrel{\R}} s_2'$.
\end{enumerate}
\end{defi}

\begin{lem}\label{lemma:up-to}
If $\R$ is a bisimulation up to $\bbisim$, then ${\R} \subseteq {\bbisim}$.
\end{lem}
\fullversion{%
\begin{proof}
It is sufficient to prove that $\mathrel{\bbisim \mathrel{\circ} \mathrel{\R}}$ is a branching bisimulation, since $\bbisim$ is reflexive.
Let $s_1\bbisim s_2\mathrel{\R} s_3$.
\begin{enumerate}
    \item Suppose $s_1\step{a}s_1'$. We distinguish two cases,
    \begin{enumerate}
        \item If $a=\tau$ and $s_1\bbisim s_1'$, then $s_1'\bbisim s_1\bbisim s_2$, so $s_1'\mathrel{\bbisim \mathrel{\circ} \mathrel{\R}}s_3$.
        \item Otherwise, we have ${a\neq\tau}\vee{s_1\not\bbisim s_1'}$. Then according to Definition~\ref{def:bbisim}, there exist $s_2''$ and $s_2'$ such that $s_2\step{}^{*}s_2''\step{a}s_2'$, $s_1\bbisim s_2''$ and $s_1'\bbisim s_2'$. Note that $s_2\bbisim s_1 \bbisim s_2''$, so by Definition~\ref{def:up-to}, there exist $s_4''$, $s_4'$ and $s_3'$ such that $s_3\step{a}s_3'$ and $s_2''\bbisim s_4'' \mathrel{\R} s_3$ and $s_2'\bbisim s_4'\mathrel{\R} s_3'$. Since $s_1'\bbisim s_2'\bbisim s_4'$ and $s_4' \mathrel{\R} s_3' $, it follows that $s_1'\mathrel{\bbisim \mathrel{\circ} \mathrel{\R}} s_3'$.

    \end{enumerate}
    \item If $s_3\step{a}s_3'$, then according to Definition~\ref{def:up-to}, there exist $s_2''$ and $s_2'$ such that $s_2\step{}^{*}s_2''\step{a}s_2'$, $s_2''\bbisim s_2$ and $s_2'\mathrel{\bbisim \mathrel{\circ} \mathrel{\R}}s_3'$. Since $s_1\bbisim s_2\bbisim s_2''$ and $s_2''\step{a}s_2'$, by Definition~\ref{def:bbisim}, there exist $s_1''$ and $s_1'$ such that $s_1\step{}^{*}s_1''\step{(a)}s_1'$ with $s_1'' \bbisim s_2''$ and $s_1'\bbisim s_2'$. Since $s_2''\bbisim s_2\mathrel{\R} s_3$ and $s_2'\mathrel{\bbisim \mathrel{\circ} \mathrel{\R}}s_3'$, it follows that $s_1''\mathrel{\bbisim \mathrel{\circ} \mathrel{\R}}s_3$ and $s_1'\mathrel{\bbisim \mathrel{\circ} \mathrel{\R}}s_3'$.
\end{enumerate}
Therefore, a branching bisimulation up to $\bbisim$ is included in $\bbisim$.
\end{proof}}


\subsection{Reactive Turing Machines and Executability}

The notion of (finite) reactive Turing machine (RTM) was put forward in \cite{BLvT2013} to mathematically characterise which behaviour is executable by a conventional computing system. In this section, we recall the definition of RTMs and the ensued notion of executable transition system. The definition of RTMs is parameterised by a set $\Atau$ of \emph{action symbols} and a set $\D$ of \emph{data symbols}. We extend $\D$ with a special symbol $\Box\notin\D$ to denote a blank tape cell; the elements of  $\Dbox=\D\cup\{\Box\}$ are called \emph{tape symbols}.
\begin{defi} \label{def:rtm}
A \emph{reactive Turing machine} (RTM) $\M$ is a tuple $(\Sta,\Dbox,\Atau,\step{},\Ins)$, where
\begin{enumerate}
\item $\Sta$ is a set of \emph{states};
\item $\Dbox$ is a set of \emph{tape symbols} including the special symbol $\Box$ denoting a blank tape cell;
  \item $\Atau$ is a set of \emph{action symbols}  including the special symbol $\tau$ denoting an unobservable event;
    \item ${\step{}}\subseteq \Sta\times\Dbox\times\Atau\times\Dbox\times\{L,R\}\times\Sta$ is a $(\Dbox\times\Atau\times\Dbox\times\{L,R\})$-labelled \emph{transition relation} (we write $s\step{a[d/e]M}t$ for $(s,d,a,e,M,t)\in{\step{}}$); and
    \item ${\Ins}\in\Sta$ is a distinguished \emph{initial state}.
    \end{enumerate}
An RTM is \emph{finite} if the sets $\Sta$, $\Dbox$ and $\Atau$ are all finite.
\end{defi}
\begin{rem}
  The reactive Turing machines proposed in \cite{BLvT2013} are finite by definition. In Section~\ref{sec:ofexecpi} we wish to investigate a relaxation of the finiteness requirement, and therefore it is convenient to provide a more general definition of the notion here. Until Section~\ref{sec:ofexecpi} all RTMs are assumed to be finite, even if we do not explicitly say so.
  \end{rem}
\begin{rem}
    The original definition of RTMs in \cite{BLvT2013} includes an extra facility to declare a subset of the states of an RTM as final states, and so does the associated notion of executable transition system. In this paper, however, our goal is to explore the relationship  between the transition systems associated with RTMs and those that can be specified in the $\pi$-calculus. Since the $\pi$-calculus does not include the facility to specify that a state has the option to terminate, we leave it out from the definition of RTMs too.
\end{rem}

Intuitively, the meaning of  a transition $s\step{a[d/e]M}t$ is that whenever $\M$ is in state $s$, and $d$ is the symbol currently read by the tape head, then it may execute the action $a$, write symbol $e$ on the tape (replacing $d$), move the read/write head one position to the left or the right on the tape (depending on whether $M=L$ or $M=R$), and then end up in state $t$.

To formalise the intuitive understanding of the operational behaviour of RTMs, we associate with every RTM $\M$ an $\Atau$-labelled transition system  $\T(\M)$. The states of $\T(\M)$ are the
configurations of $\M$, which consist of a state from $\Sta$, its tape contents, and the position of the read/write head.
We denote by $\check{\Dbox}=\{\check{d}\mid d\in\Dbox\}$ the set of \emph{marked} symbols; a \emph{tape instance} is a sequence $\delta\in(\Dbox\cup\check{\Dbox})^{*}$ such that $\delta$ contains exactly one element of $\check{\Dbox}$, indicating the position of the read/write head.
We adopt a convention to concisely denote new placement of the tape head marker. Let $\delta$ be an element of $\Dbox^{*}$. Then by $\tphdL{\delta}$ we denote the element of $(\Dbox\cup\check{\Dbox})^{*}$ obtained by placing the tape head marker on the right-most symbol of $\delta$ if $\delta$ is non-empty, and $\check{\Box}$ otherwise.
Similarly $\tphdR{\delta}$ is obtained by placing the tape head marker on the left-most symbol of $\delta$ if $\delta$ is non-empty, and $\check{\Box}$ otherwise.

\begin{defi}\label{def:lts-tm}
Let $\M=(\Sta,\Dbox,\Atau,\step{},\Ins)$ be an RTM. The \emph{transition system} $\T(\M)$ \emph{associated with} $\M$ is defined as follows:
\begin{enumerate}
\item its set of states is the set $\Conf[\M]=\{(s,\delta)\mid s\in\Sta,\ \text{$\delta$ a tape instance}\}$ of all configurations of $\M$;
    \item its transition relation ${\step{}}\subseteq{\Conf[\M]\times\Atau\times\Conf[\M]}$ is the least relation satisfying, for all $a\in\Atau,\,d,e\in\Dbox$ and $\delta_L,\delta_R\in\Dbox^{*}$:
    \begin{itemize}
        \item $(s,\delta_L\check{d}\delta_R)\step{a}(t,\tphdL{\delta_L}e\delta_R)$ iff $s\step{a[d/e]L}t$, and
        \item $(s,\delta_L\check{d}\delta_R)\step{a}(t,\delta_L e{}\tphdR{\delta_R})$ iff $s\step{a[d/e]R}t$, and
    \end{itemize}
    \item its initial state is the configuration $(\Ins,\check{\Box})$.
\end{enumerate}
\end{defi}

Turing introduced his machines to define the notion of \emph{effectively computable function} \cite{Turing1936}. By analogy, the notion of finite RTM can be used to define a notion of \emph{effectively executable behaviour}.

A transition system is \emph{(finitely) executable} if it is the transition system associated with some (finite) RTM. Usually, we shall be interested in executability up to some behavioural equivalence (e.g., the divergence-preserving or divergence-insensitive variant of branching bisimilarity).
\begin{defi}\label{def:exe}
A transition system $T$ is \emph{finitely executable up to branching bisimilarity} if there exists a finite RTM $\M$ such that $T\bbisim\T(\M)$. It is \emph{finitely executable up to divergence-preserving branching bisimilarity} if there exists a finite RTM $\M$ such that $T\bbisimd\T(\M)$.
\end{defi}

\subsection{\texorpdfstring{$\pi$}{pi}-Calculus}\label{subsec:pi}

The $\pi$-calculus was proposed by Milner, Parrow and Walker in~\cite{Milner1992} as a language to specify processes with link mobility. The expressiveness of many variants of the $\pi$-calculus has been extensively studied. In this paper, we shall consider the basic version presented in \cite{SW01}, excluding the match prefix. We recapitulate some definitions from \cite{SW01} below and refer to the book for detailed explanations.

We presuppose a countably infinite set $\N$ of names; we use strings of lower case letters for elements of $\N$.
The \emph{prefixes}, \emph{processes} and \emph{summations} of the $\pi$-calculus are, respectively, defined by the following grammar:
\begin{align*}
\pi\      & \coloneqq\ \outcap{x}{y}\ \mid\ \incap{x}{z}\ \mid\ \taucap \qquad (x,y,z\in \N)\\
P\    & \coloneqq\ M\ \mid\  P\parc P\ \mid\ \restr{z}{P}\ \mid\ \repl{P}\\
M\   & \coloneqq\ \nil\ \mid\ \pref{\pi}P \mid\ M \altc M\enskip.
\end{align*}

\newcommand{\aconv}{\mathrel{=_{\alpha}}}
\noindent
In $\pref{\incap{x}{z}}P$ and $\restr{z}{P}$, the displayed occurrence of the name $z$ is \emph{binding} with scope $P$. An occurrence of a name in a process is \emph{bound} if it is, or lies within the scope of, a binding occurrence in $P$; otherwise it is free. We use $\fn{P}$ to denote the set of names that occur free in $P$, and $\bn{P}$ to denote the set of names that occur bound in $P$. We write $P\aconv{} Q$ if $P$ and $Q$ are \emph{$\alpha$-convertible}, i.e., if $Q$ can be obtained from $P$ by a finite number of changes of bound names (see \cite{SW01} for details).

We define the operational behaviour of $\pi$-processes by means of the structural operational semantics in Table~\ref{tab:pi-semantics}, in which $\piact{}$ ranges over the set of actions of the $\pi$-calculus
\begin{equation} \label{eq:Api}
  \Api =\{\inact{x}{y},\outact{x}{y},\boutact{x}{z}\mid x,y,z \in\N\}\cup\{\tau\}
\enskip.
\end{equation}

\newcommand{\RPreftau}{\ensuremath{(\textsc{Tau})}}
\newcommand{\RPrefout}{\ensuremath{(\textsc{Out})}}
\newcommand{\RPrefin}{\ensuremath{(\textsc{Inp})}}
\newcommand{\RSuml}{\ensuremath{(\textsc{Sum-l})}}
\newcommand{\RSumr}{\ensuremath{(\textsc{Sum-r})}}
\newcommand{\RParl}{\ensuremath{(\textsc{Par-l})}}
\newcommand{\RParr}{\ensuremath{(\textsc{Par-r})}}
\newcommand{\RComl}{\ensuremath{(\textsc{Comm-l})}}
\newcommand{\RComr}{\ensuremath{(\textsc{Comm-r})}}
\newcommand{\RClosel}{\ensuremath{(\textsc{Close-l})}}
\newcommand{\RCloser}{\ensuremath{(\textsc{Close-r})}}
\newcommand{\RRes}{\ensuremath{(\textsc{Res})}}
\newcommand{\ROpen}{\ensuremath{(\textsc{Open})}}
\newcommand{\RRepact}{\ensuremath{(\textsc{Rep-act})}}
\newcommand{\RRepcomm}{\ensuremath{(\textsc{Rep-comm})}}
\newcommand{\RRepclose}{\ensuremath{(\textsc{Rep-close})}}
\newcommand{\RAlpha}{\ensuremath{(\textsc{Alpha})}}

\begin{table}[htb]
  \begin{center}
    \begin{tabular}{c} \hline \\ \\
      $\RPreftau\ \inference{\,}{\tau.P\step{\tau}P}$\quad
      $\RPrefout\ \inference{}{\overline{x}y.P\step{\overline{x}y}P}$\quad
      $\RPrefin\ \inference{}{x(y).P\step{xz}P\{z/y\}}$
      \\ \\
      $\RSuml\  \inference{P\step{\piact{}}P'}{{P+Q}\step{\piact{}}P'}$\quad
      $\RAlpha\  \inference{P\step{\piact{}}P'}{Q\step{\piact{}}P'}$\ $P\aconv Q$\\ \\
      $\RParl\ \inference{P\step{\piact{}}P'}{P\parc{Q}\step{\piact{}}P'\parc{Q}}\,\bn{\piact{}}\cap \fn{Q}=\emptyset$\quad
      $\RRepact\ \inference{P\step{\piact{}}P'}{\repl{P}\step{\piact{}}P'\parc\repl{P}}$
      \\ \\
      $\RComl\ \inference{P\step{\overline{x}y}P' & Q\step{xy}Q'}{P\parc{Q}\step{\tau}P'\parc{Q'}}$\ 
      $\RClosel\ \inference{P\step{\overline{x}(z)}P' & Q\step{xz}Q'}{P\parc{Q}\step{\tau}\restr{z}{(P'\parc{Q'})}}\,z\notin \fn{Q}$
      \\ \\
$\RRes\ \inference{P\step{\piact{}}P'}{\restr{z}{P}\step{\piact{}}\restr{z}{P'}}\,z\notin\piact{}$\quad
$\ROpen\ \inference{P\step{\overline{x}z}P'}{\restr{z}P\step{\overline{x}(z)}P'}\,z\neq x$\\ \\
      $\RRepcomm\ \inference{P\step{\outact{x}{y}}P',\,P\step{\inact{x}{y}}P''}{\repl{P}\step{\tauact{}}(P'\parc P'')\parc\repl{P}}$\quad
      $\RRepclose\ \inference{P\step{\boutact{x}{z}}P',\,P\step{\inact{x}{z}}P''}{\repl{P}\step{\tauact}\restr{z}{(P'\parc P'')}\parc\repl{P}}$ \\ \\
      \hline
  \end{tabular}
\end{center}
\caption{The operational rules for the $\pi$-calculus; the symmetric variants \RParr{}, \RComr{} and \RCloser{} of the rules \RParl{}, \RComl{} and \RClosel{}, respectively, have been omitted for conciseness.}\label{tab:pi-semantics}
\end{table}

The rules in Table~\ref{tab:pi-semantics} define on $\pi$-terms an $\Api$-labelled transition relation ${\step{}}$.
Then, we can associate with every $\pi$-term $P$ an $\Api$-labelled transition system $\T(P)=(\Sta_{P},\step{}_{P},P)$. The set of states $\Sta_{P}$ of $\T(P)$ consists of all $\pi$-terms reachable from $P$, the transition relation $\step{}_{P}$ of $\T(P)$ is obtained by restricting the transition relation $\step{}$ defined by the structural operational rules to $\Sta_{P}$ (i.e., ${\step{}_{P}}={\step{}}\cap (\Sta_{P}\times \Api \times\Sta_{P})$), and the initial state of $\T(P)$ is the $\pi$-term $P$.

For convenience, we sometimes want to abbreviate interactions that involve the transmission of no name at all, or more than one name. Instead of giving a full treatment of the polyadic $\pi$-calculus (see \cite{SW01}), we define the following abbreviations, assuming $w\not\in\fn{P}$ in both:
\begin{align*}
   \pref{\outcap{x}{\langle y_1,\dots,y_n\rangle}}P
      &\defeqn\restr{w}{\pref{\outcap{x}{w}}\pref{\outcap{w}{y_1}}\cdots\pref{\outcap{w}{y_n}}P},\ \text{and}\\
   \pref{\incap{x}{z_1,\dots,z_n}}P
       &\defeqn\pref{\incap{x}{w}}\pref{\incap{w}{z_1}}\cdots\pref{\incap{w}{z_n}}P
 \enskip.
\end{align*}

\begin{table}[htb]
  \begin{center}
    \begin{tabular}{rcl@{\qquad}rcl} \hline \\
      $P_1 + (P_2 + P_3)$ & $\scong$ & $(P_1+P_2)+P_3$
    & $P_1 \parc (P_2\parc P_3)$ & $\scong$ & $(P_1\parc P_2)\parc P_3$\\                                      
      $P_1 + P_2$ & $\scong$ & $P_2+P_1$
    & $P_1 \parc P_2$ & $\scong$ & $P_2\parc P_1$\\
      $P + \nil$ & $\scong$ & $P$
    & $P \parc \nil$ & $\scong$ & $P$\\ \\
      $\restr{z}{\restr{w}{P}}$ & $\scong$ & $\restr{w}{\restr{z}{P}}$
    & $\restr{z}{(P_1\parc P_2)}$ & $\scong$ & $P_1\parc\restr{z}{P_2}$\quad(if $z\notin\fn{P_1}$) \\
      $\restr{z}{\nil}$ & $\scong$ & $\nil$
    & $\repl{P}$ & $\scong$ & $P\parc \repl{P}$ \\ \\
      \hline
    \end{tabular}
    \end{center}
    \caption{The axioms of structural congruence.}\label{tab:scongaxioms}
  \end{table}

Divergence-preserving branching bisimilarity is not a congruence with respect to $\pi$-calculus parallel composition and restriction, due to subtle issues regarding free and bound names stemming from scope extrusion. In the remainder of this section we introduce several technical tools that we shall use in our proof that the $\pi$-calculus is behaviourally complete up to divergence-preserving branching bisimilarity in Section~\ref{sec:simulation}.

\emph{Structural congruence}, denoted by $\scong$, is the least congruence on $\pi$-terms satisfying the axioms in Table~\ref{tab:scongaxioms}. The following lemma establishes that structurally congruent $\pi$-terms are divergence-preserving branching bisimilar.
\begin{lem}\label{lemma:congbisim}
   For all  $\pi$-terms $P$ and $Q$, if $P\scong Q$, then $P\bbisimd Q$.
 \end{lem}
 \begin{proof}
   Using the Harmony Lemma \cite[Lemma 1.4.15]{SW01} it is straightforward to establish that $\scong$ is a divergence-preserving branching bisimulation, from which the lemma immediately follows.
 \end{proof}

Another useful tool in arguments establishing divergence-preserving branching bisimilarity will be the notion of deterministic internal computation.
\begin{defi}
  Let $P$ and $P'$ be $\pi$-temrs. Then $P$ has a \emph{deterministic internal computation} to $P'$ (notation: $P\dcomp P'$) if there exist $\pi$-terms $P_0,\dots,P_n$ such that $P=P_0\step{\tau}\cdots\step{\tau}P_n=P'$ and for every $\pi$-term $P_i$ ($1\leq i < n$) it holds that $P_i\step{\alpha} P_i'$ implies $\alpha=\tau$ and $P_i'=P_{i+1}$.
\end{defi}

All $\pi$-terms on a deterministic internal computation are divergence-preserving branching bisimilar, and 
\begin{lem}\label{lemma:dcompcong}
  Let $P$ and $P'$ be $\pi$-terms. If $P\dcomp P'$, then for all $z_1,\dots,z_n\in\N$ and for all $\pi$-terms $Q$ it holds that $\restr{z_1,\dots,z_n}{(P\parc Q)}\bbisimd\restr{z_1,\dots,z_n}{(P'\parc Q)}$.
\end{lem}

\noindent
We say that a $\pi$-term $P$ has \emph{reachable bound output} if there exist $\pi$-terms $P_0,\dots,P_n,P'$, actions $\alpha_1,\dots,\alpha_n\in\Api$ and names $x,z\in\N$ such that
\begin{equation*}
  P=P_0\step{\alpha_1}\cdots\step{\alpha_n}P_n\step{\boutact{x}{z}}P'\enskip.
\end{equation*}

\begin{lem}\label{lemma:noboutcong}
  Let $P_1$, $P_2$, $Q_1$ and $Q_2$ be $\pi$-terms without reachable bound output. If $P_1\bbisimd Q_1$ and $P_2\bbisimd Q_2$, then $P_1\parc P_2\bbisimd Q_1\parc Q_2$.
\end{lem}

\noindent
Let $P$ be a $\pi$-term and let $z\in\N$. We say that $P$ \emph{eventually outputs $z$}  if there exist $\pi$-terms $P_0,\dots,P_n,P'$, actions $\alpha_1,\dots,\alpha_n\in\Api$ and a name $x\in\N$ such that
\begin{equation*}
  P=P_0\step{\alpha_1}\cdots\step{\alpha_n}P_n\step{\outact{x}{z}}P'\enskip.
\end{equation*}
\begin{lem}\label{lemma:restrcong}
  Let $P$ and $Q$, and let $z\in\N$ be such that neither $P$ nor $Q$ eventually outputs $z$. If $P\bbisimd Q$, then $\restr{z}{P}\bbisimd\restr{z}{Q}$.
\end{lem}

\section{The \texorpdfstring{$\pi$}{pi}-Calculus is Behaviourally Complete}\label{sec:simulation}

In the previous section, we have introduced the $\pi$-calculus as a language to specify behaviour of systems with link mobility, and we have proposed RTMs to define a notion of executable behaviour. In this section we prove that every executable behaviour can be specified in the $\pi$-calculus up to divergence-preserving branching bisimilarity. To this end, we associate with every RTM $\M$ a $\pi$-term $P$ that simulates the behaviour of $\M$ up to divergence-preserving branching bisimilarity, that is, $\T(\M)\bbisim^{\Delta}\T(P)$.

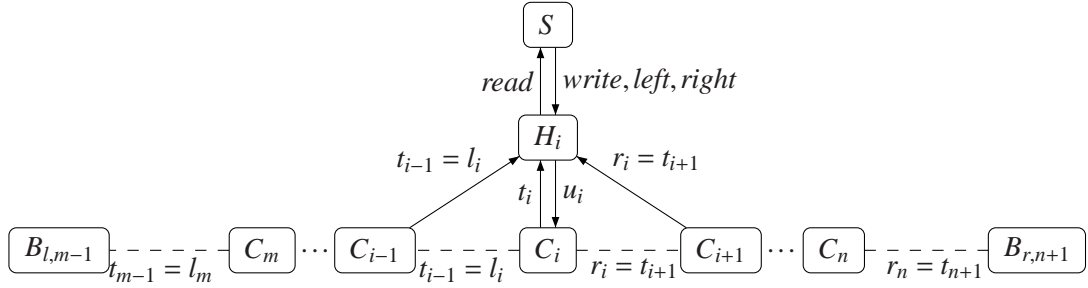
\begin{figure}[htb]
\begin{center}
\begin{tikzpicture}[-,>=stealth']

   \node[state] (Blm-1) {$B_{l,m-1}$};
   \node[state,right of=Blm-1,node distance=3.0cm] (Cm) {$C_m$};
   \node[right of=Cm,node distance=0.75cm] (dots1) {$\dots$};
  \node[state,right of=dots1,node distance=0.7cm] (Ci-1) {$C_{i-1}$};
  \node[state,right of=Ci-1,node distance=2.6cm] (Ci) {$C_i$};
  \node[state,right of=Ci,node distance=2.6cm] (Ci+1) {$C_{i+1}$};
     \node[right of=Ci+1,node distance=0.75cm] (dots2) {$\dots$};
  \node[state,right of=dots2,node distance=0.7cm] (Cn) {$C_n$};
  \node[state,right of=Cn,node distance=3.0cm] (Brn+1) {$B_{r,m+1}$};
  \node[state,above of=Ci,node distance=1.5cm] (Hi) {$H_i$};
  \node[state,above of=Hi,node distance=1.5cm] (S) {$S$};
 
  \path
    (Blm-1) edge[dashed]  node[anchor=south,above]{$t_{m-1}=l_m$} (Cm)
    (Ci-1) edge[dashed]  node[anchor=south,above]{$t_{i-1}=l_i$} (Ci)
    (Ci) edge[dashed] node[anchor=south,above]{$r_i=t_{i+1}$} (Ci+1)
    (Cn) edge[dashed]  node[anchor=south,above]{$r_{n}=t_{m+1}$} (Brn+1)
    (Ci) edge[->,bend left] node[left]{$t_i$} (Hi)
    (Hi) edge[->,bend left] node[left]{$\textit{read}$} (S)
    (S) edge[->,bend left] node[right]{$\textit{write},\textit{left},\textit{right}$} (Hi)
    (Hi) edge[->,bend left] node[right]{$u_i$} (Ci)
    (Ci-1.north) edge[->] node[left, near end,xshift=-.2cm]{$t_{i-1}=l_i$} (Hi.west)
    (Ci+1.north) edge[->] node[right, near end,xshift=.2cm]{$r_i=t_{i+1}$} (Hi.east)
   ;

\end{tikzpicture}
\caption{\label{fig:specification}Specification of an RTM utilizing the linking structure of the $\pi$-calculus}
\end{center}
\end{figure}

The structure of our specification is illustrated in Figure~\ref{fig:specification}. In this figure, each node represents a parallel component of the specification, each labelled arrow stands for a communication channel, with the arrow pointing from sender to receiver. The dashed lines represent links between cells that are achieved by instantiating parameters with the same name. The equalities on arrows and dashed lines indicate identifications that will thus be made; for instance, the equality $t_{i-1}=l_i$ indicates that the parameter $t$ of the process $C_{i-1}$ will be instantiated with the same name as the parameter $l$ of the process $C_i$. The specification consists of a generic finite specification of the behaviour of a tape (parallel components $H_k$, $B_{l,k}$, $C_k$, $B_{r,k}$ in Figure~\ref{fig:specification}), and a finite specification of a control process that is specific for the RTM $\M$ under consideration (parallel component $S$ in Figure~\ref{fig:specification}). We first discuss the generic specification of the tape in Section~\ref{subsec:tape}, then we discuss how to add a suitable control process specific for $\M$ in Section~\ref{subsec:fincontrol} proving that $\M$ is simulated by the parallel composition of the two parts.

\subsection{Tape} \label{subsec:tape}

In~\cite{BBK1987}, the behaviour of the tape of a Turing machine is finitely specified in ACP$_{\tau}$ making use of finite specifications of two stacks. That specification is not easily modified to take intermediate termination into account, and therefore, in~\cite{BLvT2013}, an alternative solution is presented, specifying the behaviour of a tape in TCP$_{\tau}$ by using a finite specification of a queue (see also \cite{BBR2010}). In this paper, we will exploit the link passing feature of the $\pi$-calculus to give a more direct specification. In particular, we shall model the tape as a collection of cells endowed with a link structure that organises them in a linear fashion.

We first give an informal description of the behaviour of a tape. The state of a tape is characterised by a tape instance $\delta_L\check{d}\delta_R$, consisting of a finite (but unbounded) sequence of data with the current position of the tape head indicated by ~$\check{}$~. The tape may then exhibit the following observable actions:
\begin{enumerate}
    \item $\overline{\textit{read}}\,d$: the datum under the tape head is output along the channel $\textit{read}$;
    \item $\textit{write}(e)$: a datum $e$ is written on the position of the tape head, resulting in a new tape instance $\delta_L\check{e}\delta_R$; and
    \item $\textit{left}$, $\textit{right}$:  the tape head moves one position left or right, resulting in ${\tphdL{\delta_L}}d\delta_R$ or $\delta_L d \tphdR{\delta_R}$, respectively.
\end{enumerate}

Henceforth, we assume that tape symbols are included in the set of names, i.e., that $\Dbox\subseteq\N$.

In our $\pi$-calculus specification of the behaviour of a tape, each individual tape cell is specified as a separate component, and there is a separate component modelling the tape head. A tape cell stores a datum $d$, represented by a free name in the specification, and it has pointers $l$ and $r$ to its left and right neighbour cells. Furthermore, it has two links to  the component modelling the tape head: the link $u$ is used by the tape head for updating the datum, and the link $t$ serves as a general communication channel for communicating all relevant information about the cell to the tape head.

The following $\pi$-term represents the behaviour of a tape cell:
\begin{eqnarray*}
C &\ddef& c(t,l,r,u,d).C(t,l,r,u,d)\\
C(t,l,r,u,d) &\ddef& u(e).\overline{c}\langle t,l,r,u,e\rangle.\nil+\overline{t}\langle l,r,u,d\rangle.\overline{c}\langle t,l,r,u,d\rangle.\nil
\enskip.
\end{eqnarray*}
A cell is created by a synchronisation on name $c$, by which all relevant information about the cell is passed; we shall have a component $!C$ facilitate the generation of new incarnations of existing tape cells. Note that the behaviour of an individual tape cell $C(t,l,r,u,d)$ is as follows: either it receives along channel $u$ an update $e$ for its datum $d$, after which it recreates itself with datum $e$ in place of $d$; or it outputs all relevant information about itself (i.e., the links to its left and right neighbours, its update channel $u$, and
the stored datum $d$) to the tape head along channel $t$, after which it recreates itself. The following lemma, which is a straightforward consequence of the definition of $C$ and the operational rules of the $\pi$-calculus, expresses that recreation proceeds via a deterministic internal computation.

\begin{lem}\label{lemma:recrC}
$\restr{c}{(\overline{c}\langle t,l,r,u,d\rangle.\nil\parc{!C})}\dcomp\scong \restr{c}{(C(t,l,r,u,d)\parc{!C})}$
\end{lem}

At any moment, the number of tape cells will be finite.  To model the unbounded nature of the tape, we define a process $B$ that serves to generate new blank tape cells on either side of the  tape whenever needed:
\begin{eqnarray*}
B &\ddef&b_l(t,r).\restr{u,l}{B_l(t,l,r,u)} + b_r(t,l).\restr{u,r}{B_r(t,l,r,u)}\\
B_l(t,l,r,u) &\ddef& \overline{t}\langle l,r,u,\Box\rangle.\overline{b_l}\langle l,t\rangle.\overline{c}\langle t,l,r,u,\Box\rangle.\nil \\
B_r(t,l,r,u) &\ddef& \overline{t}\langle l,r,u,\Box\rangle. \overline{b_r}\langle r,t\rangle.\overline{c}\langle t,l,r,u,\Box\rangle.\nil
\enskip.
\end{eqnarray*}

Note that $B$ offers the choice to either create a blank tape cell at the left-hand side of the tape through $B_l(t,l,r,u)$, or a blank tape cell at the right-hand side of the tape through $B_r(t,l,r,u)$. In the first case, suppose the original leftmost cell has the channels $t_o$ and $l_o$, for itself and its left neighbour, respectively, then for the new cell, we have $t=l_o$ and $r=t_o$, in order to maintain the links to its neighbour. Moreover, at the creation of the new blank cell, two new links are created too: $u$ is the update channel of the new blank cell, and $l$ will later be used as the link to generate another cell. Thus, an extra blank cell is generated on the left through $\overline{b_l}\langle l,t\rangle.\nil$ and the original blank cell on the left is promoted to a regular cell by $\overline{c}\langle t,l,r,u,\Box\rangle.\nil$. In the second case, a symmetrical procedure is implemented by $B_r(t,l,r,u)$. 

\begin{lem}\label{lemma:recrB}
  We have
  \begin{multline*}
    \restr{b_l,b_r,c}{(\overline{b_l}\langle l,t\rangle.\overline{c}\langle t,l,r,u,\Box\rangle.\nil\parc\repl{B}\parc\repl{C})} \\ \dcomp\scong \restr{b_l,b_r,c,u',l'}{(B_l(l,l',t,u')\parc C(t,l,r,u,\Box)\parc\repl{B}\parc\repl{C})}
  \end{multline*}
  and
  \begin{multline*}
    \restr{b_l,b_r,c}{(\overline{b_r}\langle r,t\rangle.\overline{c}\langle t,l,r,u,\Box\rangle.\nil\parc\repl{B}\parc\repl{C})} \\ \dcomp\scong \restr{b_l,b_r,c,u',r'}{(B_r(r,t,r',u')\parc C(t,l,r,u,\Box)\parc\repl{B}\parc\repl{C})}
  \end{multline*}
\end{lem}

Throughout the simulation of an RTM, the number of parallel components modelling individual tape cells will grow. We shall presuppose a numbering of these parallel components with consecutive integers from some interval $[m,n]$ ($m$ and $n$ are integers such that $m\leq n$), in agreement with the link structure. The numbering is reflected by a naming scheme that adds the subscript $i$ to the links $t$, $l$, $r$, $u$ and $d$ of the $i$th cell ($m\leq i \leq n$). We abbreviate $C(t_i,l_i,r_i,u_i,d_i)$ by $C_i(d_i)$, and $B_l(t_i,l_i,r_i,u_i)$ and $B_l(t_i,l_i,r_i,u_i)$ by $B_{l,i}$ and $B_{r,i}$, respectively. Let $\vec{d}_{[m,n]}=d_m,d_{m+1},\dots,d_{n-1},d_n$; we define:
\begin{equation*}
  \textit{Cells}_{[m,n]}(\vec{d}_{[m,n]}) \ddef \restr{b_l, b_r, c}{(B_{l,m-1} \parc{  C_m(d_m)}
    \parc{ \cdots}
    \parc{ C_n(d_n) }\parc{ B_{r,n+1}} \parc {! C} \parc {! B})}
\enskip.
\end{equation*}

The component modelling the tape head serves as the interface between the tape cells and the RTM-specific control process.
It is defined as:
\begin{eqnarray*}
H &\ddef& h(t,l,r,u,d).H(t,l,r,u,d)\\
H(t,l,r,u,d)&\ddef&\overline{\textit{read}}\,d.\overline{h}\langle t,l,r,u,d\rangle.\nil+\textit{write}(e).\overline{u}\, e.\overline{h}\langle t,l,r,u,e\rangle.\nil\\
 && \quad\mbox{} + \textit{left}.l(l',r',u',d').\overline{h}\langle l,l',r',u',d'\rangle.\nil\\
 && \quad\mbox{} + \textit{right}.r(l',r',u',d').\overline{h}\langle r,l',r',u',d'\rangle.\nil
 \enskip.
\end{eqnarray*}

\noindent
The tape head maintains two links to the current cell (a communication channel $t$ and an update channel $u$), as well as links to its left and right neighbour cells ($l$ and $r$, respectively). Furthermore, the tape head remembers the datum $d$ in the current cell. The datum $d$ may be output along the $\textit{read}$-channel. Furthermore, a new datum $e$ may be received  through the $\textit{write}$-channel, which is then forwarded through the update channel $u$ to the current cell. Finally, the tape head may receive instructions to move left or right, which has the effect of receiving information about the left or right neighbours of the current cell through $l$ or $r$, respectively. In all cases, a new incarnation of the tape head is started, with a call on the $h$-channel.

\begin{lem}\label{lemma:recrH}
$\restr{h}{(\overline{h}\langle t,l,r,u,d\rangle.\nil\parc{!H})}\dcomp\scong \restr{h}{(H(t,l,r,u,d)\parc{!H})}$
\end{lem}

\noindent
Let
$\vec{t}_{[m,n]}  =t_m,t_{m+1},\dots,t_{n-1},t_n$,
let
$\vec{u}_{[m,n]}  =u_m,u_{m+1},\dots,u_{n-1},u_n$, and
let
  $H_i(d_i)  =H(t_i,l_i,r_i,u_i,d_i)$;
we define
\begin{equation*}
  \textit{Tape}_{[m,n]}^i(\vec{d}_{[m,n]}) \ddef \restr{\vec{t}_{[m-1,n+1]}, \vec{u}_{[m,n]}}{(\restr{h}{(H_i(d_i)\parc {!H})} \parc{ \textit{Cells}_{[m,n]}(\vec{d}_{[m,n]})})}
\enskip.
\end{equation*}

We shall write $P\step{a}\bbisimd P'$ for ``there is a $P''$ such that $P\step{a}P''$ and $P''\bbisimd P'$''.

\begin{lem}\label{lemma:tape-behaviour}
    $\textit{Tape}_{[m,n]}^i(\vec{d}_{[m,n]})\step{\alpha}T'$ if, and only if, at least one of the following holds:
\begin{enumerate}
\item $\alpha=\outact{\textit{read}}{d_i}$ and $T' \bbisimd \textit{Tape}_{[m,n]}^i(\vec{d}_{[m,n]})$, or
\item $\alpha=\inact{\textit{write}}{e}$ and $T' \bbisimd \textit{Tape}_{[m,n]}^i(d_{[m,i-1]},e,d_{[i+1,n]})$, or
\item $\alpha=\inact{\textit{left}}{}$, $i>m$ and $T'\bbisimd \textit{Tape}_{[m,n]}^{i-1}(\vec{d}_{[m,n]})$, or
\item $\alpha=\inact{\textit{left}}{}$, $i=m$ and $T'\bbisimd\textit{Tape}_{[m-1,n]}^{i-1}(\Box,\vec{d}_{[m,n]})$, or
\item $\alpha=\inact{\textit{right}}{}$, $i<n$ and $T'\bbisimd\textit{Tape}_{[m,n]}^{i+1}(\vec{d}_{[m,n]})$, or
\item $\alpha=\inact{\textit{right}}{}$, $i=n$ and $T'\bbisimd \textit{Tape}_{[m,n+1]}^{i+1}(\vec{d}_{[m,n]},\Box)$.
\end{enumerate}
\end{lem}
\fullversion{%
  \begin{proof}
    The component $\textit{Cells}_{[m,n]}(\vec{d}_{[m,n]})$ of $\textit{Tape}_{[m,n]}^i(\vec{d}_{[m,n]})$ only admits interactions on the channels $\vec{t}_{[m-1,n+1]}$ and $\vec{u}_{[m,n]}$, which are restricted in $\textit{Tape}_{[m,n]}^i(\vec{d}_{[m,n]})$. Hence, the only transitions afforded by $\textit{Tape}_{[m,n]}^i(\vec{d}_{[m,n]})$ are those faciliated by the component $H_i(d_i)$, from which it is clear that $\alpha=\outact{\textit{read}}{d_i}$, $\alpha=\inact{\textit{write}}{e}$, $\alpha=\inact{\textit{left}}{}$, or $\alpha=\inact{\textit{right}}{}$. After each of these transitions, the component $H_i(d_i)$ initiates deterministic internal computations of one or more parallel components:
  
  If $\alpha=\outact{\textit{read}}{d_i}$, then, note that, up to structural congruence, $T'$ is obtained from $\textit{Tape}_{[m,n]}^i(\vec{d}_{[m,n]})$ by replacing the component $H_i(d_i)$ by $\overline{h}\langle t_i,l_i,r_i,u_i,d_i\rangle.\nil$. Since
  \begin{equation*}
    \restr{h}{(\overline{h}\langle t_i,l_i,r_i,u_i,d_i\rangle.\nil\parc{!H})}\dcomp\scong \restr{h}{(H(t_i,l_i,r_i,u_i,d_i)\parc{!H})}\enskip,
  \end{equation*}
  by Lemma~\ref{lemma:recrH}, it follows by Lemma~\ref{lemma:dcompcong} that $T'\bbisimd\textit{Tape}_{[m,n]}^i(\vec{d}_{[m,n]})$.

  If $\alpha=\inact{\textit{write}}{e}$, then, note that, up to structural congruence, $T'$ is obtained from $\textit{Tape}_{[m,n]}^i(\vec{d}_{[m,n]})$ by replacing the component $H_i(d_i)$ by $\overline{u}\, e.\overline{h}\langle t_i,l_i,r_i,u_i,e\rangle.\nil$. Then, $\overline{u}\, e$ interacts with $\textit{Cells}_{[m,n]}(\vec{d}_{[m,n]})$ to update $d_i$ to $e$, giving rise to a deterministic internal communication $T' \dcomp T''$. Moreover, $T''$ is obtained from $\textit{Tape}_{[m,n]}^i(d_{[m,i-1]},e,d_{[i+1,n]})$ by replacing the component $H_i(e)$ by $\overline{h}\langle t_i,l_i,r_i,u_i,e\rangle.\nil$ and the component $C_i(d_i)$ by $\overline{c}\langle t_i,l_i,r_i,u_i,e\rangle.\nil$. Since, by Lemma~\ref{lemma:recrC},
  \begin{equation*}
    \restr{c}{(\overline{c}\langle t_i,l_i,r_i,u_i,e\rangle.\nil\parc{!C})}\dcomp\scong \restr{c}{(C(t_i,l_i,r_i,u_i,e)\parc{!C})}\enskip,
  \end{equation*}
  and, by Lemma~\ref{lemma:recrH},
  \begin{equation*}
    \restr{h}{(\overline{h}\langle t_i,l_i,r_i,u_i,e\rangle.\nil\parc{!H})}\dcomp\scong \restr{h}{(H(t_i,l_i,r_i,u_i,e)\parc{!H})}\enskip,
  \end{equation*}
  it follows by Lemma~\ref{lemma:dcompcong} that
    $T''\bbisimd\textit{Tape}_{[m,n]}^i(d_{[m,i-1]},e,d_{[i+1,n]})$,
    and hence
    \begin{equation*}
      T''\bbisimd \textit{Tape}_{[m,n]}^i(d_{[m,i-1]},e,d_{[i+1,n]})
      \enskip.
     \end{equation*}

    If $\alpha=\inact{\textit{left}}{}$ or $\alpha=\inact{\textit{right}}{}$ then the arguments are similar as above, using also Lemma~\ref{lemma:recrB} if $i=m$ or $i=n$, respectively.
\end{proof}}

\subsection{Finite control}\label{subsec:fincontrol}

We associate with every RTM $\M=(\Sta_{\M},\Dbox,\Atau,\step{}_{\M},\Ins_{\M})$ a finite specification of its control process. Here $m$ can be either $\textit{left}$ or $\textit{right}$.
\begin{eqnarray}
S&\ddef& \sum_{s\in \Sta_{\M}}s.\sum_{d\in\D_{\Box}}d.S_{s,d}\nonumber\\
S_{s,d}&\ddef&\sum_{(s,d,a,e,m,t)\in\step{}_{\M}}a.\overline{\textit{write}}\,e.\overline{m}.\textit{read}(f).\overline{t}.\overline{f}.\nil\nonumber
\end{eqnarray}
We define
\begin{equation*}
  \textit{Control}_{s,d} \ddef S_{s,d}\parc{!S}
\enskip.
\end{equation*}
The following lemma characterises the behaviour of the control process.
\begin{lem}\label{lemma:transition-step}
Let $\M=(\Sta_{\M},\Dbox,\Atau,\step{}_{\M},\Ins_{\M})$ be an RTM. Then
 \begin{equation*}
\textit{Control}_{s,d}\step{a}\step{\overline{\textit{write}}\,e}\step{\overline{m}}\step{\textit{read}\,f} \dcomp\scong\textit{Control}_{t,f}
\enskip.
\end{equation*}
if and only if $(s,d,a,e,m,t)\in\step{}_{\M}$.
\end{lem}

Given an RTM $\M$, we associate with every configuration $(s,\delta_L\check{d}\delta_R)$ a $\pi$-term $M_{s,\delta_L\check{d}\delta_R}$, consisting of a parallel composition of the specifications of its tape instance and control process. Let $\vec{s}=s_1,s_2,\ldots,s_m\in \Sta_{\M}$, let $\vec{e}=e_1,e_2,\ldots,e_n\in \D_{\Box}$, and let $\vec{r}=\textit{read},\textit{write},\textit{left},\textit{right}$; we define
 \begin{equation*}
  M_{s,\delta_L\check{d}\delta_R}=\restr{\vec{e},\vec{s},\vec{r}}{(\textit{Control}_{s,d}\parc{ \textit{Tape}^{i}_{[m,n]}(\vec{d}_{[m,n]})})},\,where\, \vec{d}_{[m,n]}=\delta_L\check{d}\delta_R
\enskip.
\end{equation*}
 The following lemma establishes that $M_{s,\delta_L\check{d}\delta_R}$ simulates the execution steps of the RTM in the configuration $(s,\delta_L\check{d}\delta_R)$.
 \begin{lem}\label{lemma:behaviour-m}
 Given an RTM $\M=(\Sta_{\M},\Dbox,\Atau,\step{}_{\M},\Ins_{\M})$, for every configuration $(s,\delta_L\check{d}\delta_R)$, we have
   \begin{equation*}
   M_{s,\delta_L\check{d}\delta_R}\step{a}\bbisimd M_{t,\delta_L'\check{f}\delta_R'}
   \end{equation*}
   if and only if there is a transition $(s,\delta_L\check{d}\delta_R)\step{a}(t,\delta_L'\check{f}\delta_R')$.
 \end{lem}
\fullversion{%
  \begin{proof}
    On the one hand, if $M_{s,\delta_L\check{d}\delta_R}\step{a}\bbisimd M_{t,\delta_L'\check{f}\delta_R'}$, then
    \begin{equation*}
      \textit{Control}_{s,d}\step{a}\step{\overline{\textit{write}}\,e}\step{\overline{m}}\step{\textit{read}\,f} \dcomp\scong\textit{Control}_{t,f}\enskip,
    \end{equation*}
    so by Lemma~\ref{lemma:transition-step}, $(s,d,a,e,m,t)\in\step{}_{\M}$.

    On the other hand, if $(s,\delta_L\check{d}\delta_R)\step{a}(t,\delta_L'\check{f}\delta_R')$, then $(s,d,a,e,m,t)\in\step{}_{\M}$.
    Hence, by Lemma~\ref{lemma:transition-step}, we have
\begin{equation*}
M_{s,\delta_L\check{d}\delta_R}\step{a}\restr{\vec{e},\vec{s},\vec{r}}{(\overline{\textit{write}}\,e.\overline{m}.\textit{read}(f).\overline{t}.\overline{f}.\nil\parc{!S}\parc{\textit{Tape}^{i}_{[m,n]}(\vec{d}_{[m,n]})})}=M'
\enskip.
\end{equation*}
It then remains to prove that $M'\bbisimd M_{t,\delta_L'\check{f}\delta_R'}$.

To this end, first note that by Lemma~\ref{lemma:tape-behaviour}, we get
  \begin{equation*}
  M'\dcomp\scong \restr{\vec{e},\vec{s},\vec{r}}{(\overline{m}.\textit{read}(f).\overline{t}.\overline{f}.\nil\parc{!S}\parc{T'})}
   \enskip,
  \end{equation*}
  where $T'\bbisimd \textit{Tape}^{i}_{[m,n]}(d_m,\ldots,d_{i-1}, e,d_{i+1},\ldots,d_n)$, so by Lemmas~\ref{lemma:congbisim}, \ref{lemma:dcompcong}, \ref{lemma:noboutcong} and \ref{lemma:restrcong}
  \begin{multline*}
    M'\bbisimd \\ \restr{\vec{e},\vec{s},\vec{r}}{(\overline{m}.\textit{read}(f).\overline{t}.\overline{f}.\nil\parc{!S}\parc \textit{Tape}^{i}_{[m,n]}(d_m,\ldots,d_{i-1}, e,d_{i+1},\ldots,d_n))}=M''\enskip.
  \end{multline*}
  Then, again by Lemma~\ref{lemma:tape-behaviour}, we get
   \begin{equation*}
  M''\dcomp\scong \restr{\vec{e},\vec{s},\vec{r}}{(\textit{read}(f).\overline{t}.\overline{f}.\nil\parc{!S}\parc{T''})}
   \enskip,
  \end{equation*}
  where $T''\bbisimd \textit{Tape}^{j}_{[m,n]}(d_m,\ldots,d_{i-1}, e,d_{i+1},\ldots,d_n)$ and $j=i-1$ if $m=\textit{left}$ and $j=i+1$ if $m=\textit{right}$.
  So, by Lemmas~\ref{lemma:congbisim}, \ref{lemma:dcompcong}, \ref{lemma:noboutcong} and \ref{lemma:restrcong},
  \begin{multline*}
    M''\bbisimd \\ \restr{\vec{e},\vec{s},\vec{r}}{(\textit{read}(f).\overline{t}.\overline{f}.\nil\parc{!S}\parc \textit{Tape}^{j}_{[m,n]}(d_m,\ldots,d_{i-1}, e,d_{i+1},\ldots,d_n))}=M'''\enskip.
  \end{multline*}
  Then, with another application of Lemma~\ref{lemma:tape-behaviour}, we get
   \begin{equation*}
  M'''\dcomp\scong \restr{\vec{e},\vec{s},\vec{r}}{(\overline{t}.\overline{f}.\nil\parc{!S}\parc{T'''})}
   \enskip,
  \end{equation*}
  where $T'''\bbisimd \textit{Tape}^{j}_{[m,n]}(d_m,\ldots,d_{i-1}, e,d_{i+1},\ldots,d_n)$.
  So, by Lemmas~\ref{lemma:congbisim}, \ref{lemma:dcompcong}, \ref{lemma:noboutcong} and \ref{lemma:restrcong},
  \begin{equation*}
    M'''\bbisimd \restr{\vec{e},\vec{s},\vec{r}}{(\overline{t}.\overline{f}.\nil\parc{!S}\parc \textit{Tape}^{j}_{[m,n]}(d_m,\ldots,d_{i-1}, e,d_{i+1},\ldots,d_n))}=M''''\enskip.
  \end{equation*}
  And finally,
   \begin{equation*}
  M''''\dcomp\scong M_{t,\delta_L'\check{f}\delta_R'}
   \enskip,
 \end{equation*}
 where $\delta_L'f\delta_R'=d_m,\ldots,d_{i-1}, e,d_{i+1},\ldots,d_n$.
 \end{proof}}

\begin{thm}\label{thm:rtm-pi-spec}
  For every RTM $\M$, we have that $\T(M_{\Ins,\check{\Box}})\bbisimd\T(\M)$.
\end{thm}
\fullversion{%
\begin{proof}
  Let $\M=(\Sta,\Dbox,\Atau,\step{}',\Ins)$, let $i\notin\Atau$, and let $\M'=(\Sta,\Dbox,\Atau,\step{}',\Ins)$ be the RTM obtained from $\M$ by replacing all $\tau$-transitions by $i$-transitions. Clearly, $\T(\M)$ is isomorphic to $\T(\M')$ up to a renaming of $i$ to $\tau$. Hence, in order to prove that $\T(M_{\Ins,\check{\Box}})\bbisimd\T(\M)$, it suffices to prove that $\T(M_{\Ins,\check{\Box}}')\bbisimd\T(\M')$, where $M_{\Ins,\check{\Box}}'$ is the $\pi$-term associated with $(\Ins,\check{\Box})$ given $\M'$.
  
Using Lemma~\ref{lemma:behaviour-m} it is straightforward to establish that the relation
\begin{equation*}
\R'=
\{(M_{s,\delta_L\check{d}\delta_R}',(s,\delta_L\check{d}\delta_R))\mid s\in \Sta,\,\delta_L,\,\delta_R\in \Dbox^{*},\,\check{d}\in\check{\Dbox}\}
\end{equation*}
is a branching bisimulation up to $\bbisim$.
So, by Lemma~\ref{lemma:up-to}, it follows that $\T(M_{\Ins,\check{\Box}}')\bbisim\T(\M')$.

It remains to show that divergence is preserved too.

Note that there is no $\tau$-transition in $\M'$, which means $\T(\M')$ has no divergence. Then, by Lemma~\ref{lemma:behaviour-m}, the specification of a certain configuration $M_{s,\delta_L\check{d}\delta_R}'$ can only do $a$-labelled transitions, where $a\in\A\cup \{i\}$, i.e.
\begin{equation*}
   M_{s,\delta_L\check{d}\delta_R}'\step{a}M'\bbisimd M_{t,\delta_L'\check{f}\delta_R'}'
   \enskip.
\end{equation*}
Since there is no $\tau$ transition from the term $M_{t,\delta_L'\check{f}\delta_R'}'$, it follows that $M'$ has no divergence either. Hence, no $\pi$-term reachable from $M_{s,\delta_L\check{d}\delta_R}'$ admits a divergence, and therefore we have,

\begin{equation*}
\T(M_{\Ins,\check{\Box}}')\bbisimd\T(\M')
\enskip.
\end{equation*}

Finally, we switch back to $\M$, by changing all the $i$ labelled transition to $\tau$, and we let $M_{\Ins,\check{\Box}}$ be the specification of the initial state of $\M$. We can also establish that the relation
\begin{equation*}
\R=
\{(M_{s,\delta_L\check{d}\delta_R},(s,\delta_L\check{d}\delta_R))\mid s\in \Sta,\,\delta_L,\,\delta_R\in \Dbox^{*},\,\check{d}\in\check{\Dbox}\}
\enskip.
\end{equation*}
is a branching bisimulation up to $\bbisim$. Moreover, note that every infinite sequence of the form $\step{i}\step{}^{*}\step{i}\step{}^{*}\cdots$ from $M_{\Ins,\check{\Box}}'$ corresponds with a infinite sequence of the form $\step{i}\step{i}\cdots$ from $\M'$, and vice versa. Additionally, there is no divergence from $M_{\Ins,\check{\Box}}'$. Therefore, every infinite $\tau$-labelled transition sequence from $M_{\Ins,\check{\Box}}$ corresponds with an infinite $\tau$-labelled transition sequence from $\M$. We conclude that ${\R}\subseteq{\bbisimd}$.
\end{proof}}

Thus we have the following expressiveness result for the $\pi$-calculus.
\begin{cor}\label{coro:pi-exp}
The $\pi$-calculus is behaviourally complete up to divergence-preserving branching bisimilarity.
\end{cor}

\section{On the Executability of the \texorpdfstring{$\pi$}{pi}-Calculus} \label{sec:ofexecpi}

We have proved that every executable behaviour can be specified in the $\pi$-calculus modulo divergence-preserving branching bisimilarity. We shall now investigate to what extent behaviour specified in the $\pi$-calculus is executable. Recall that we have defined executable behaviour as behaviour of an RTM. So, in order to prove that the behaviour specified by a $\pi$-term is executable, we need to show that the transition system associated with this $\pi$-term is behaviourally equivalent to the transition system associated with some RTM.

Note, however, that there is a mismatch between the formalisms of RTMs and $\pi$-calculus. On the one hand, the notion of  RTM as we have defined it in Section~\ref{subsec:behaviour} presupposes \emph{finite} sets $\Atau$ and $\Dbox$ of actions and data symbols, and also the transition relation of an RTM is \emph{finite}. 
As a consequence, the transition system associated with an RTM is finitely branching, and, in fact, its branching degree is bounded by a natural number. (Note that this does not mean that RTMs cannot deal with data of unbounded size; it only means that it has to be encoded using finitely many symbols.)
The $\pi$-calculus, on the other hand, presupposes an infinite set of names by which an infinite set of actions $\Api$ is generated. Furthermore, the transition system associated with a $\pi$-term by the structural operational semantics (see Table~\ref{tab:pi-semantics}) may contain states with an infinite branching degree, even modulo branching bisimilarity, due to the rules for input prefix and bound output prefix.

From the assumption that the set of actions and the transition relation of an RTM must be finite it immediately follows that there are $\pi$-terms that cannot be simulated. The following example illustrates that it is not enough to relax just those two requirements.
\begin{exa}~\label{exa:piinfbranch}
  Consider the $\pi$-term $P=\pref{\incap{x}{y}}\pref{\outcap{y}{y}}\nil$. According to $\RPrefin$ (see Table~\ref{tab:pi-semantics}), $P$ affords, for every $z\in\N$, a transition $P\step{xz}\pref{\outcap{z}{z}}\nil$, and since $\pref{\outcap{z}{z}}\nil\step{\outcap{z}{z}}\nil$ is the only transition of $\pref{\outcap{z}{z}}\nil$, it follows that $\pref{\outcap{z'}{z'}}\nil\not\bbisim\pref{\outcap{z''}{z''}}\nil$ if $z'\neq z''$.

  Now suppose that $\M=(\Sta,\Dbox,\Atau,\step{},\Ins)$ is an RTM, except that we allow its set of actions to be the infinite set $\Api$ and also we allow its transition relation to be infinite, and assume that $\T(\M)\bbisim\T(P)$.

  Let $C=(\Ins,\check{\Box})$ be the initial configuration of $\M$. We have $C\bbisim P$, so $C$ affords, for every $z\in\N$, a transition sequence
  \begin{equation*}
    C\step{}^{*}\step{\inact{x}{z}}\step{}^{*}C_z\step{\outcap{z}{z}}C_z'\enskip,
  \end{equation*}
  with $C_z\bbisim\pref{\outcap{z}{z}}\nil$ and $C_z'\bbisim\nil$.

The transition rules of RTMs are of the form $s\step{a[d/e]M}t$, where $s,\,t\in\Sta$, and $d,\,e\in\Dbox$; we call the pair $(s,d)$ the trigger of this rule. A configuration $(s',\delta_L\check{d'}\delta_R)$ satisfies the trigger $(s,d)$ if $s=s'$ and $d=d'$. Now observe that a rule $s\step{a[d/e]M}s'$ gives
rise to an $a$-transition from every configuration satisfying its trigger $(s,d)$. Since $\Sta$ and $\Dbox$ are finite, there are finitely many triggers.

So, in the infinite collection of configurations $C_z$ ($z\in\N$), there are at least two configurations, say $C_{z'}$ and $C_{z''}$ with $z'\neq z''$, satisfying the same trigger $(s,d)$; these configurations must have the same outgoing transitions. It follows that $C_{z'}\step{\outcap{z''}{z''}}C_{z'}''$, which contradicts $C_{z'}\bbisim\pref{\outcap{z'}{z'}}\nil$. We conclude that $\T(\M)\not\bbisim\T(P)$.
\end{exa}

The example illustrates that it is not enough, for the simulation modulo branching bisimilarity of $\pi$-terms, to relax the finiteness restriction on the set of actions and the transition relation. We should also allow the set of states $\Sta$ or the set of data sybols $\Dbox$ to be infinite. Simply lifting the finiteness requirement on either yields a notion of RTM that is arguably too expressive: in both cases, every countable transition system can be simulated up to divergence-preserving branching bisimilarity \cite[Section 6.1]{Yan18}. The transition system associated with a $\pi$-term is clearly countable if the set of names $\N$ is countable.

Following the work of Boja\'nczyk, Klin, Lasota, and Toru\'nczyk \cite{BKLT13}, we consider in this section a more modest relaxation of the finiteness requirements on RTMs. We propose \emph{orbit-finite} RTMs and an associated notion of \emph{orbit-finite executability}, and then we prove that every $\pi$-term can be simulated up to branching bisimilarity by an orbit-finite RTM. 

\subsection{Orbit-Finite Executability}

The notion of orbit-finite executability that we introduce below is based on the notion of \emph{orbit-finite set} proposed in \cite{BKL11}. Below, we briefly recap the definitions; we refer to \cite{Boj19} for an extensive treatment and elaborate explanations.

\newcommand{\Sym}[1]{\ensuremath{\mathit{Sym}(#1)}} \newcommand{\gact}{\ensuremath{\mathbin{\cdot}}}
\newcommand{\neutral}{\ensuremath{\mathalpha{id}}}
\newcommand{\orbeq}{\ensuremath{\mathrel{\sim}}}
Let $\Atoms$ be a countably infinite set of \emph{atoms}. We denote by $\Sym{\Atoms}$ the group of all permutations on $\Atoms$, and denote by $\neutral$ the neutral element of $\Sym{\Atoms}$. An \emph{action} of $\Sym{\Atoms}$ on a set $X$ is a binary operation
  $\gact:\Sym{\Atoms}\times X\rightarrow X$
  such that $\neutral\gact x=x$ for all $x\in X$ and $\pi\rho\gact x=\pi\gact(\rho\gact x)$ for all $\pi,\rho\in\Sym{\Atoms}$ and $x\in X$. The symbol $\gact$ will often be omitted. For the sets with atoms appearing in the remainder of this paper, $\gact$ will always be specified in the expected manner: the elements are denoted by expressions involving atoms and, for every element $x$ it is assumed that $\pi\gact x$ is obtained by replacing every atom $a$ occurring in the expression denoting $x$ by $\pi(a)$.

A set $A\subseteq\Atoms$ \emph{supports} an element $x\in X$ if for all $\pi\in\Sym{\Atoms}$ such that $\pi(a)=a$ for all $a\in A$ it holds that $\pi x =x$. If for all $x\in X$ there exists a finite set $A_x\subseteq\Atoms$ that supports $x$, then $X$ is called a \emph{nominal set}.

An action of $\Sym{\Atoms}$ on $X$ induces an equivalence relation $\orbeq$ on $X$, defined by $x\orbeq y$ if, and only if, there exists $\pi\in\Sym{\Atoms}$ such that $\pi\gact x= y$. The equivalence relation $\orbeq$ partitions $X$ into equivalence classes called \emph{orbits}, and $X$ is \emph{orbit-finite} if $\orbeq$ partitions $X$ into finitely many orbits.

\begin{defi}
  A reactive Turing machine $(\Sta,\Dbox,\Atau,\step{},\Ins)$ is \emph{orbit-finite} if $\Sta$, $\Dbox$, $\Atau$ and $\step{}$ are \emph{orbit-finite} nominal sets. A transition system is \emph{orbit-finitely executable} if it is the transition system associated with some orbit-finite reactive Turing machine.
\end{defi}

\subsection{The \texorpdfstring{$\pi$}{pi}-Calculus is Orbit-Finitely Executable}

\newcommand{\Sim}[1]{\ensuremath{\mathtt{Sim}(#1)}}
\newcommand{\Init}[1]{\ensuremath{\mathtt{Init}(#1)}}
\newcommand{\Gen}{\ensuremath{\mathtt{Gen}}}
\newcommand{\Chck}{\ensuremath{\mathtt{Check}}}
\newcommand{\Exec}{\ensuremath{\mathtt{Exec}}}
   
To prove that the $\pi$-calculus is orbit-finitely executable up to branching bisimilarity, we should establish that for every $\pi$-term $P_0$ there exists an orbit-finite RTM $\Sim{P_0}$ such that $\T(P_0)\bbisim\T(\Sim{P_0})$. The RTM $\Sim{P_0}$ will be orbit-finite with respect to a set of atoms
$\Atoms$ that consists of the set $\N$ of names of the $\pi$-calculus, i.e., we assume
\begin{equation*}
  \Atoms=\N
  \enskip.
\end{equation*}
We shall define $\Sim{P_0}$ as the union of several smaller RTMs that take care of specific aspects of the simulation; we shall refer to these smaller RTMs as \emph{fragments} of $\Sim{P_0}$. Before we proceed to describe these fragments in detail, we first give a broad overview of the fragments we need. In the overview we  assume that $\pi$-terms, transitions and derivations of transitions can be stored on the tape of the RTM; this will also be discussed in more detail below.
\begin{description}
\item[\normalfont\itshape Initialise] By design, an RTM starts with an empty tape, while for the simulation it is convenient to assume that (an encoding of) the $\pi$-term representing the current state is written on tape. The purpose of the \emph{initialise} fragment $\Init{P_0}$ is to write an encoding of $P_0$, the $\pi$-term to be simulated, on tape.
\item[\normalfont \itshape Generate transition] It is assumed that the \emph{generate} fragment $\Gen$ starts executing with the encoding of an arbitrary $\pi$-term $P$ stored on tape. Its purpose is then to generate (the encoding of) a transition that is derivable according to the operational semantics of the $\pi$-calculus and has $P$ as source. This is achieved by non-deterministically writing an arbitrary sequence of symbols on tape, and then verifying whether the sequence encodes a derivation of a transition and whether this transition has $P$ as source. If so, then the \emph{generate transition} fragment erases everything from the tape except for the encoding of the derived transition and proceeds with the \emph{execute transition} fragment described next; if not, then it erases the generated sequence of symbols and restarts the transition generation fragment with the same $\pi$-term on tape.
\item[\normalfont \itshape Execute transition] It is assumed that the \emph{execute} fragment $\Exec{}$ starts executing with the encoding of a transition $P\step{\alpha}P'$ written on the tape. Its purpose is to execute the associated action and thereafer leave only the target $P'$ of the transition on the tape, returning to the \emph{generate transition} fragment.
\end{description}
Note that only the $\Init{P_0}$ fragment will be specific for every $\pi$-term $P_0$. The fragments $\Gen$ and $\Exec$ are generic; their behaviour depends on their initial tape contents. We proceed to first define the sets $\D$ and $\A$ associated with $\Sim{P_0}$, and then discuss the definitions of the three fragments in more detail.

\newcommand{\opsym}{\ensuremath{\overline{\phantom{x}}}}
\newcommand{\pisym}{\ensuremath{\D_{\pi}}}
\newcommand{\auxsym}{\ensuremath{\D_{\textit{aux}}}}
\newcommand{\encoding}[1]{\ensuremath{\lceil{#1}\rceil}}

\subsubsection*{Orbit-finite sets of data and action symbols}

We let the set of data symbols $\D$ of $\Sim{P_0}$ consist of the set of $\pi$-calculus names $\N$, extended with two \emph{finite} sets of symbols $\pisym$ and $\auxsym$. To store a $\pi$-term on the tape of an RTM, we use the symbols in $\N$ and 
  \begin{equation*}\label{eq:pitapesymbols}
     \pisym=\{ \opsym{}, {(}, {)}, \tau, \nil, {.}, +, \mathalpha{\mid}, \nu, \mathalpha{!} \}\enskip,
   \end{equation*}
writing $\overline{x}$ on the tape as $\opsym{}x$.
   The symbols in $\auxsym$ will serve as auxiliary symbols in computations; we assume that$\auxsym$ at least includes the symbols $\#$, $\langle$, $\rangle$, ${-}$, ${>}$, $\therefore$, $[$, and $]$.
   The sets $\pisym$ and $\auxsym$ are assumed to be mutually disjoint and also disjoint from $\N$, and they are assumed to have been constructed using standard set-theoretic methods, not involving atoms. We have
   \begin{equation*}
     \D = \pisym\cup\N\cup\auxsym \enskip.
   \end{equation*}
   The orbits of $\D$ are $\N$ and all the singleton subsets of $\pisym$ and $\auxsym$, so $\D$ has $|\pisym|+1+|\auxsym|$ orbits. Hence $\D$ and clearly also $\Dbox$ are orbit-finite.
   Trivially, every $\pi$-term $P$ can be specified with a sequence of symbols from $\D$; for clarity of presentation, we shall distinguish a $\pi$-term $P$ from its specification as a sequence of symbols from $\D$ writing $\encoding{P}$ for the latter.

   The set of action symbols $\Atau$ of $\Sim{P_0}$ will be the set of all $\pi$-calculus actions given the set of names $\N$ (see Equation~\ref{eq:Api}), i.e.,
   \begin{equation*}
     \Atau=\{\inact{x}{y},\outact{x}{y},\boutact{x}{z}\mid x,y,z \in\N\}\cup\{\tau\}\enskip.
   \end{equation*}
   Note that $\inact{x_1}{y_1}\orbeq\inact{x_2}{y_2}$ if, and only if, either $x_1=y_1$ and $x_2=y_2$, or $x_1\neq y_1$ and $x_2\neq y_2$. Hence, the set $\{\inact{x}{y}\mid x,y\in\N\}$ has two orbits, and, by similar reasoning, so do $\{\outact{x}{y}\mid x,y\in\N\}$ and $\{\boutact{x}{z}\mid x,z\in\N\}$. It follows that $\Atau$ has seven orbits in total, and hence is orbit-finite. Clearly, every $\pi$-calculus action $\alpha$ can be specified on tape using the symbols in $\D$; we denote this sequence by $\encoding{\alpha}$.

   \subsubsection*{Initialise}

   The initialise fragment $\Init{P_0}=(\Sta_{\Init{P_0}},\Dbox,\Atau,\step{}_{\Init{P_0}},\Ins_{\Init{P_0}})$ should simply write the encoding $\encoding{P_0}$ of $P_0$ on the tape. This is straightforward: Let $n=|\encoding{P_0}|$, and assume $\encoding{P_0}=d_0\cdots d_n$. We define $\Sta_{\Init{P_0}}=\{\Ins_{\Init{P_0}},s_0,\dots,s_{n},\Ins_{\Gen}\}$. The execution of $\Init{P_0}$ starts with a transition
   \begin{equation*}
     \Ins_{\Init{P_0}}\step{\tau[\Box/d_0]R}_{\Init{P_0}} s_0\enskip,
   \end{equation*}
   has for every $0\leq i < n$ a transition of the form
   \begin{equation*}
     s_i\step{\tau[\Box/d_i]R}_{\Init{P_0}} s_{i+1} \enskip,
   \end{equation*}
   and ends with a transition
   \begin{equation*}
     s_{n}\step{\tau[\Box/\#]R}_{\Init{P_0}}\Ins_{\Gen}\enskip.
   \end{equation*}
   The latter transition marks the end of the sequence $\encoding{P_0}$ on the tape with the symbol $\#$, and its target is the initial state $\Ins_{\Gen}$ of the generate fragment $\Gen$.
   Since both $\Sta_{\Init{P_0}}$ and $\step{}_{\Init{P_0}}$ are finite and $\Dbox$ and $\Atau$ are orbit-finite, it follows that $\Init{P_0}$ is an orbit-finite RTM.

   \subsubsection*{Generate transition}
   The generate transition fragment $\Gen=(S_{\Gen},\Dbox,\Atau,\step{}_{\Gen},\Ins_{\Gen})$ starts with the sequence of symbols $\encoding{P}\#$ written on tape and the tape head positioned on the first blank immediately to the right of the sequence. The purpose of the fragment is to compute a transition $P\step{\alpha}P'$, specified on tape as a sequence $\langle \encoding{P}{-}\encoding{\alpha}{>}\encoding{P'}\rangle$. To this end, $\Gen$ will non-deterministically generate a sequence of symbols in $\D\backslash\{\#\}$ followed by $\#$, and subsequently checks whether the sequence between the two $\#$ symbols on tape encodes a derivation of a transition of the form $P\step{\alpha}P'$.

   Generating a sequence of symbols is straightforward: it requires only two states $\Ins_{\Gen}$ and $\Ins_{\Chck}$ and transitions
   \begin{equation*}
     \Ins_{\Gen}\step{\tau[\Box/d]R}_{\Gen}\Ins_{\Gen}\qquad (d\in\D\backslash\{\#\})
   \end{equation*}
   and
   \begin{equation*}
     \Ins_{\Gen}\step{\tau[\Box/\#]R}_{\Gen}\Ins_{\Chck}\enskip.
   \end{equation*}
   Note that, since $\D$ is orbit-finite and hence also $\D\backslash\{\#\}$ is orbit-finite, there are orbit-finitely many transitions of the first type. So, this part of $\Gen$ is clearly orbit-finite.

   The verification procedure, which starts in $\Ins_{\Chck}$ is tedious, but straightforward given the operational semantics. We shall refrain from specifying it in detail and only comment on the subprocedures that need to be carried out by this fragment, and argue that they are orbit-finite. First, however, we need to explain how a derivation can be (uniquely) encoded as a sequence of symbols from $\D-\{\#\}$. We adopt the convention that (sub)derivations are included in square brackets and the conclusion of a (sub)deriviation is preceded with the symbol $\therefore$. For instance, the derivation
   \begin{equation*} \inference[\RRes{}]{\inference[\RComl{}]{\inference[\RPrefout]{}{\pref{\outcap{x}{z}}\nil\step{\outact{x}{z}}\nil}&\inference[\RPrefin]{}{\pref{\incap{x}{y}}\nil\step{\inact{x}{z}}\nil}}{{\pref{\outcap{x}{z}}\nil\parc\pref{\incap{x}{y}}\nil}\step{\tau}{\nil\parc\nil}}}{\restr{z}{(\pref{\outcap{x}{z}}\nil\parc\pref{\incap{x}{y}}\nil)}\step{\tau}{\restr{z}{(\nil\parc\nil)}}}
   \end{equation*}
   is specified on tape as
  \begin{multline*}
     [[[{\therefore{}}{\langle}\opsym{}xz{.}\nil{-}\opsym{}xz{>}\nil{\rangle}]
       [{\therefore{}}{\langle}\incap{x}{y}{.}\nil{-}xz{>}\nil{\rangle}]
       {\therefore{}}{\langle}\opsym{}xz{.}\nil{\parc}\incap{x}{y}{.}\nil{-}\tau{>}\nil{\parc}\nil{\rangle}] \\
       {\therefore{}}{\langle}(\nu z){(\opsym{}xz{.}\nil{\parc}\pref{\incap{x}{y}}\nil)}{-}\tau{>}(\nu z)(\nil{\parc}\nil){\rangle}]
       \enskip.
     \end{multline*}%
   To verify whether a sequence of symbols from $\D\backslash\{\#\}$ indeed encodes a derivation, $\Gen$ implements a procedure consisting of the following steps:
   \begin{enumerate}
   \item\label{item:step1}
     Check whether the occurrences of the symbols $[$, $\therefore$ and $]$ between the two occurrences of the symbol $\#$ on the tape represent a tree structure. If so, then continue to the next step; otherwise, remove all symbols to the right of the left-most occurrence of the symbol $\#$ and return to $\Ins_{\Gen}$.
   \item\label{item:step2}
     Check whether all sequences of symbols between subsequent occurrences of $\therefore$ and $]$ encode transitions. If so, then continue to the next step; otherwise, remove all symbols to the right of the left-most occurrence of the symbol $\#$ and return to $\Ins_{\Gen}$.
   \item\label{item:step3}
     Check whether all individual inferences of the represented derivation are instances of a rule of the operational semantics (see Table~\ref{tab:pi-semantics}). If so, then continue to the next step; otherwise, remove all symbols to the right of the left-most occurrence of the symbol $\#$ and return to $\Ins_{\Gen}$.
   \item\label{item:step4}
     Check whether the sequence of symbols that constitutes the source of the right-most transition, i.e., the conclusion of the derivation specified on tape, matches the sequence $\encoding{P}$ preceding the left-most occurrence of $\#$ on the tape. If so, then erase everything except the transition $\langle \encoding{P}{-}\encoding{\alpha}{>}\encoding{P'}\rangle$ from the tape and transition to $\Ins_{\Exec}$. Otherwise, remove all symbols to the right of the left-most occurrence of the symbol $\#$ and return to $\Ins_{\Gen}$.
   \end{enumerate}
   The computations described in first two steps of the procedure can be done without any special consideration of individual names. The parts of $\Gen$ implementing these steps have finitely many states and orbit-finitely many transitions. Step~\ref{item:step4} can be implemented as a simple comparison procedure: the sequence of symbols to the left of the left-most $\#$ must be compared to the sequence of symbols representing the source of the right-most transition. Such a comparison can be done by comparing one symbol at a time, but it will involve remembering that symbol in a state. Hence this part of $\Gen$ will require an infinite set of states and an infinite set of transitions, which can, however, both be partitioned into finitely many $\pisym\cup\N$-indexed orbits.

   The implementation of step~\ref{item:step3} is computationally more involved. We discuss, for each of the operational rules in Table~\ref{tab:pi-semantics} (see p.~\pageref{tab:pi-semantics}), what needs to be done:
   \begin{description}
   \item[\normalfont\RPreftau{}] It needs to be checked that there are no premises (i.e., immediately left of the $\therefore$ symbol there is the $[$ symbol). Furthermore, it needs to be checked that the source of the transition is a $\tau$ prefix  (i.e., starts with the symbols $\tau$ and $.$), that the label of the transition is $\tau$ (i.e., the symbols $-$ and $>$ there is only the symbol $\tau$), and that the operand of the prefix is equal to the target of the transition (i.e., the sequence of symbols between the symbol $.$ and the symbol $-$ is identical to the sequence of symbols between $>$ and $\rangle$). 
   \item[\normalfont\RPrefout{}] It needs to be checked that there are no premises. Furthermore, it needs to be checked that the source of the transition is of the form $\pref{\outcap{x}{y}}P$, that the label of the transition is $\outact{x}{y}$ and that the target of the transition is $P$.
   \item[\normalfont\RPrefin{}] It needs to be checked that there are no premises. Futhermore, it needs to be checked that the source of the transition is of the form $\pref{\incap{x}{y}}P$, that the label of the transition is $\inact{x}{z}$ for some arbitrary name $z\in\N$, and that the target is obtained from $P$ by substituting all free occurrences of $y$ in $P$ by $z$. The latter operation can, e.g., be carried out by remembering the pair $(y,z)$ in states and carrying out a symbol-wise comparison in which the symbol $y$ must be matched by $z$ rather than $y$, except while in the scope of a $y$-binding construct $\incap{x}{y}$ or $\restr{y}{}$. The part of $\Gen$ that takes care of the comparison has finitely many $(y,z)$-indexed orbits.
     \item[\normalfont\RSuml{}] It needs to be checked that there is one premise, that the source of the conclusion is of the form $P+Q$, that $P$ is the source of the premise, that the labels of the premise and the conclusion are identical, and so are the targets of the source and the premise.
     \item[\normalfont\RSumr{}] Analogous to $\RSuml{}$.
     \item[\normalfont\RAlpha{}] It needs to be checked that there is one premise, that the target of the premise is identical to the target of the conclusion, and that the source of the premise and the source of the conclusion are $\alpha$-convertible. The latter could, e.g., be implemented as a variation on the comparison procedure with a renaming operation built-in  (cf.\ also \cite[Section 4]{Pit16}). Whenever, during the comparison of $\encoding{P}$ and $\encoding{Q}$, a binder, say $\incap{x}{y_1}$ in $\encoding{P}$ and $\incap{x}{y_2}$ in $\encoding{Q}$ or $\restr{y_1}{}$ in $\encoding{P}$ and $\restr{y_2}{}$ in $\encoding{Q}$, is encountered, then
       \begin{enumerate}
       \item a name $z$ is determined that is fresh for both the remainders of $\encoding{P}$ and $\encoding{Q}$,
       \item all occurrences of $y_1$ in the remainder of $\encoding{P}$ and all occurences of $y_2$ in the remainder of $\encoding{Q}$ are replaced by $z$, and
       \item the comparison continues.
       \end{enumerate}
       Note that generating a name that is fresh with respect to some sequence of symbols on tape can be achieved by non-deterministically writing an arbitrary name on tape and subsequently checking whether the name already occurs in the sequence or not. If it does occur, then the name is not fresh for the sequence and the procedure must be repeated. If it does not occur, then the name is fresh for the sequence.
     \item[\normalfont\RParl] It needs to be checked that there is one premise. Furthermore, it needs to be checked that the source of the conclusion is of the form $P\parc Q$, that $P$ is the source of the premise, that the labels of the premise and the conclusion are identical, and that the target of the conclusion is the parallel composition of the target of the premise and $Q$. Finally, it should be checked that the side condition $\bn{\alpha}\cap\fn{Q}=\emptyset$ is satisfied. To this end, a sequence of all free names occurring in $Q$ should be compiled and then it should be checked whether the bound name of $\alpha$ does not occur in that sequence.
     \item[\normalfont\RParr] Analogous to $\RParl{}$.
     \item[\normalfont\RComl] It needs to be checked that there are two premises. Futhermore, it needs to be checked that the source of the conclusion is of the form $P\parc Q$, that $P$ is the source of the first premise, that $Q$ is the source of the second premise, and that the target of the conclusion is the parallel composition of the targets of the premises. Finally, it needs to be checked that the label of the first premise is $\outact{x}{y}$ and the label of the second premise is $\inact{x}{y}$ for some names $x,y\in\N$, and that the label of the conclusion is $\tau$.
     \item[\normalfont\RComr] Analogous to $\RComl$.
     \item[\normalfont\RClosel] It needs to be checked that there are two premises. Furthermore, it needs to be checked that the source of the conclusion is of the form $P\parc Q$, that $P$ is the source of the first premise, that $Q$ is the source of the second premise, and that the target of the conclusion is of the form $\restr{z}{(P'\parc Q')}$, where $P'$ and $Q'$ are the targets of the premises. It also needs to be checked that the label of the first premise is $\boutact{x}{z}$, for some $x\in \N$, that the label of the second premise is $\inact{x}{z}$, and that the label of the conclusion is $\tau$. Finaly, it should be checked that $z\notin\fn{Q}$, by first compiling the sequence of names with a free occurrence in $Q$ and then checking whether $z$ appears in the sequence.
     \item[\normalfont\RCloser] Analogous to $\RClosel$.
     \item[\normalfont\RRes] It needs to be checked that there is one premise. Furthermore, it needs to be checked that the source of the conclusion is of the form $\restr{z}{P}$, that $P$ is the source of the premise, that the target is $\restr{z}{P'}$ where $P'$ is the target of the conclusion, that the labels of the premise and the conclusion are identical, and that the name $z$ does not appear in $\alpha$.
     \item[\normalfont\ROpen] It needs to be checked that there is one premise. Furthermore, it needs to be checked that the source of the conclusion is of the form $\restr{z}{P}$, that $P$ is the source of the premise, and that the targets of the premise and the conclusion are identical. Finally, it needs to be checked that the label of the premise is $\outact{x}{z}$ for some names $x\neq z$, and that the label of the conclusion then is $\boutact{x}{z}$.
     \item[\normalfont\RRepcomm] It needs to be checked that there are two premises. Furthermore, it needs to be checked that the source of the conclusion is of the form $\repl{P}$, that the source of both premises is $P$, and that the target of the conclusion is the parallel composition of, on the one hand, the parallel composition of the targets of the premises and, on the other hand, $\repl{P}$. Finally, it needs to be checked that the label of the first premise is $\outact{x}{y}$ for some names $x,y\in\N$, that the label of the second premise is then $\inact{x}{y}$, and that the label of the conclusion is $\tau$.
     \item[\normalfont\RRepclose] It needs to be checked that there are two premises. Furthermore, it needs to be checked that the source of the conclusion is of the form $\repl{P}$, that the source of both premises is $P$, and that the target of the conclusion is a restriction $\restr{z}{}$ applied to a parallel composition of, on the one hand, the parallel composition of the targets of the premises and, on the other hand, $\repl{P}$. Finally, it needs to be checked that the label of the first premise is $\boutact{x}{z}$ for some name $x\in\N$, that the label of the second premise is $\inact{x}{z}$, and that the label of the conclusion is $\tau$.
     \end{description}

     In step~\ref{item:step3}, $\Gen$ applies, for every individual inference, the procedures associated above with the operational rules of the $\pi$-calculus one after the other until either one of them succeeds or it has been determined that none of them succeeds. In the first case, all symbols to the right of the left-most occurrence of the symbol $\#$ are removed and $\Gen$ returns to $\Ins_{\Gen}$. In the second case, $\Gen$ proceeds with the next individual inference.

   All transitions associated with $\Gen$ are deemed internal, i.e., are labelled with $\tau$. Moreover, from every state of $\Gen$ at least one of $\Ins_{\Gen}$ and $\Ins_{\Exec}$ is reachable.

   \subsubsection*{Execute transition}

   The execute transition fragment $\Exec=(\Sta_{\Exec},\Dbox,\Atau,\step{}_{\Exec},\Ins_{\Exec})$ starts with the encoding of a transition
     $\langle\encoding{P}{-}\encoding{\alpha}{>}\encoding{P'}\rangle$
     written on the tape; we assume that the tape head is positioned on the first symbol of $\encoding{\alpha}$.
     The initial state $\Ins_{\Exec}$ admits the following transition sequences:
     \begin{gather*}
       \Ins_{\Exec} \step{\tau[\tau/\tau]R}_{\Exec} s_{\tau}\step{\tau[{>}/{>}]R}_{\Exec} \mathit{cnt}\enskip,\\
       \Ins_{\Exec} \step{\tau[x/x]R}_{\Exec} s^x_{\mathit{in}} \step{\tau[y/y]R}_{\Exec} s^{x,y}_{\mathit{in}} \step{\inact{x}{y}[{>}/{>}]R}_{\Exec} \mathit{cnt} \qquad(x,y\in\N)\enskip,\ \text{and}\\
       \Ins_{\Exec} \step{\tau[\opsym{}/\opsym{}]R}_{\Exec} s_{\mathrm{out}} \step{\tau[x/x]R}_{\Exec} s^{x}_{\mathrm{out}}\qquad(x\in\N)\enskip.
     \end{gather*}
     From the states  $s^x_{\mathit{out}}$ we need to distinguish two possible continuations, depending on whether $\alpha$ represents a regular or a bound output:
     \begin{equation*}
       s^x_{\mathit{out}} \step{\tau[y/y]R}_{\Exec} s^{x,y}_{\mathit{out}}\step{\outact{x}{y}[{>}/{>}]R}_{\Exec} \mathit{cnt}\qquad(x,y\in\N)\enskip,
     \end{equation*}
     and
     \begin{multline*}
       s^x_{\mathit{out}} \step{\tau[{(}/{(}]R}_{\Exec} t^x \step{\tau[z/z]R}_{\Exec}t^{x,z}\step{\tau[{)}/{)}R]}_{\Exec} s^{x,z}_{\mathit{bout}} \\ \mbox{} \step{\boutact{x}{z}[{>}/{>}]R}_{\Exec} \mathit{cnt}\qquad (x,z\in\N)\enskip.
     \end{multline*}
     In the state $\mathit{cnt}$ an internal (i.e., $\tau$-labelled) deterministic \emph{continue} procedure is started which ensures that the contents of the tape is $\encoding{P'}\#$. The last transition of this procedure leads to $\Ins_{\Gen}$ with the tape head positioned on the first blank immediately to the right of the sequence.

     Furthermore, for every $s\in\{s_{\tau}\}\cup\{s^{x,y}_{\mathit{in}},s^{x,y}_{\mathit{out}}\mid x,y\in\N\}\cup\{s^{x,z}_{\mathit{bout}}\mid x,z\in\N\}$, $\Exec$ includes a transition
     \begin{equation*}
       s \step{\tau[{>}/{>}]R}_{\Exec} \mathit{abrt} \enskip.
     \end{equation*}
     In the state $\mathit{abrt}$ an internal (i.e., $\tau$-labelled) deterministic \emph{abort} procedure is started which serves to restore the contents of the tape to $\encoding{P}\#$ leading to the state $\Ins_{\Gen}$ with the tape head positioned on the first blank immediately to the right of the sequence. It is necessary to have this abort procedure to ensure that the (non-deterministic) choice for this particular transition is made upon its actual execution and not before. At the same time, note that the abort procedure introduces divergence into the simulation: $\Ins_{\Gen}$ affords a non-empty sequence of $\tau$-transitions to $\mathit{abrt}$, which, in turn, affords a non-empty sequence of $\tau$-transitions back to $\Ins_{\Gen}$.

     From the descriptions above, which are parameterised in at most two names, it is clear that the part of $\Exec{}$ transitions leading up to the action execution is orbit-finite. Furthermore, the continue and abort procedures do not give any special treatment to individual elements of $\N$ (names are either skipped or erased), so these procedures require finitely many states and orbit-finitely many transitions.

     We now define $\Sim{P_0}$ as the union of the three fragments $\Init{P_0}$, $\Gen$ and $\Exec$, i.e.,
     \begin{equation*}
       \Sim{P_0} = (
         \Sta_{\Init{P_0}}\cup\Sta_{\Gen}\cup\Sta_{\Exec},
         \Dbox,
         \Atau,
         {\step{}_{\Init{P_0}}}\cup{\step{}_{\Gen}}\cup{\step{}_{\Exec}},
         \Ins_{\Init{P_0}}
      )\enskip.
    \end{equation*}
    (We assume that $\Ins_{\Gen}$ is shared between all three fragments and $\Ins_{Exec}$ is shared between the fragmens $\Gen$ and $\Exec$, but otherwise the sets of states of the three fragments are disjoint.)
    We now have the following theorem.

\begin{thm}
   For every $\pi$-term $P$ there exists an orbit-finite reactive Turing machine $\Sim{P_0}$ such that $\T(P_0)\bbisim\T(\Sim{P_0})$.
 \end{thm}
 \begin{proof}
   We have already argued that for every $P_0$ there exists an orbit-finite RTM $\Sim{P_0}$.
   It remains to establish a branching bisimulation $\mathcal{R}$ from $\T(P_0)$ to $\T(\Sim{P_0})$ such that $P_0\mathrel{\mathcal{R}}(\Ins_{\Init{P_0}}, \check{\Box})$.

   We first introduce the following notations
   \begin{enumerate}
   \item $\Init{P_0,\check{\Box}}$ denotes the set of all configurations reachable from $(\Ins_{\Init{P_0}},\check{\Box})$;
 \item for every $P$ reachable from $P_0$ in $\T(P_0)$, $\Gen(P)$ denotes the set of all configurations reachable from $(\Ins_{\Gen},\encoding{P}\#\check{\Box})$ in $\T(\Sim{P_0})$.
 \item for every $P$ reachable from $P_0$ in $\T(P_0)$, $\Exec(P)$ denotes the set of all configurations that are either on a path from $(\Ins_{\Exec},\langle\encoding{P}{-}\encoding{\check{\alpha}}{>}\encoding{P'})$ to a configuration $(s,\langle\encoding{P}{-}\encoding{\alpha}\check{>}\rangle)$ (with $s\in \{s_{\tau}\}\cup\{s^{x,y}_{\mathit{in}},s^{x,y}_{\mathit{out}}\mid x,y\in\N\}\cup\{s^{x,z}_{\mathit{bout}}\mid x,z\in\N\}$) or on a path from $(\mathit{abrt},\langle\encoding{P}{-}\encoding{\check{\alpha}}{>}\encoding{\check{P'}}\rangle)$ to $(\Ins_{\Gen},\encoding{P}\#\check{\Box})$; and
   \item for every $P'$ reachable from $P_0$ in $\T(P_0)$, $\Exec'(P')$ denotes the set of all configurations that are on a path from $(\mathit{cnt},\langle\encoding{P}{-}\encoding{\alpha}{>}\encoding{\check{P'}}\rangle$ to $(\Ins_{\Gen},\encoding{P'}\#\check{\Box})$.   
   \end{enumerate}
   We can now define the relation $\mathcal{R}$ by
   \begin{multline*}
     \mathcal{R} =
     \{(P_0,c)\mid c\in \Init{P_0,\check{\Box}}\}\cup\mbox{}\\
     \qquad\qquad\{(P,c)\mid \text{$P$ reachable from $P_0$ in $\T(P_0)$},\ c\in\Gen(P)\cup\Exec(P)\}\cup\mbox{}\\
     \{(P',c)\mid \text{$P'$ reachable from $P_0$ in $\T(P_0)$},\ c\in\Exec'(P')\}
     \enskip.
   \end{multline*}
   From $(\Ins_{\Init{P_0}},\check{\Box})$ there is a deterministic internal computation leading to $(\Ins_{\Gen},\encoding{P}\#\check{\Box})$ and $P_0$ is related according to $\mathcal{R}$ to all intermediate states of this internal computation.
   The fragments $\Gen$ and $\Exec$ have been designed such that all configurations in $\Gen(P)$ and $\Exec(P)$ are reachable from the configuration $(\Ins_{\Gen},\encoding{P}\#\check{\Box})$ by $\tau$-transitions and, moreover, from all those configurations the configuration $(\Ins_{\Gen},\encoding{P}\#\check{\Box})$ is reachable by $\tau$-transitions. It follows that $\Gen(P)\cup\Exec(P)$ is a so-called $\tau$-cluster: every configuration is reachable from every other configuration and the only exits from the cluster are the transitions of the form
   \begin{multline*}
     (s,\langle\encoding{P}{-}\encoding{\alpha}\check{>}\encoding{P'}\rangle)
     \step{\alpha}
     (c,\langle\encoding{P}{-}\encoding{\alpha}{>}\encoding{\check{P'}}\rangle)
     \\  (s\in \{s_{\tau}\}\cup\{s^{x,y}_{\mathit{in}},s^{x,y}_{\mathit{out}}\mid x,y\in\N\}\cup\{s^{x,z}_{\mathit{bout}}\mid x,z\in\N\})      \enskip.
   \end{multline*}
   Note that these transitions simulate and are simulated by transitions $P\step{\alpha}P'$.
   Finally, we have that from $(c,\langle\encoding{P}{-}\encoding{\alpha}{>}\encoding{\check{P'}}\rangle)$ there is a deterministic internal computation leading to $(\Ins_{\Gen},\encoding{P'}\#\check{\Box})$, which concludes the argument that $\mathcal{R}$ is a branching bisimulation.
 \end{proof}

 \begin{cor}
   The $\pi$-calculus is orbit-finitely executable modulo branching bisimilarity.
 \end{cor}

 \section{Conclusions} \label{sec:conclusion}

Milner already established in~\cite{Milner1992} that the $\pi$-calculus is \emph{computationally complete}, by exhibiting an encoding of the $\lambda$-calculus in the $\pi$-calculus by which every reduction in the $\lambda$-calculus is simulated by a sequence of reductions in the $\pi$-calculus. We have established that the $\pi$-calculus is \emph{behaviourally complete} up to divergence-preserving branching bisimilarity, which is the finest reasonable notion of behavioural equivalence \cite{Glabbeek1993}. This implies that the $\pi$-calculus is also behaviourally complete up to the weaker notions of behavioural equivalence usually used in the context of the $\pi$-calculus \cite{SW01}.
Interestingly, the proof does not rely on recursion and the finite specification of a queue, as does the proof in \cite{BLvT2013} that the process calculus TCP$_\tau$ is behaviourally complete. Instead, the specification of finite RTMs in the $\pi$-calculus uses replication and link mobility to directly specify a tape process. An alternative specification of finite RTMs in a process calculus without recursion is presented in \cite{BLY17}, which, instead of replication, uses iteration and nesting operators \cite{BBP94,BP01}, and also relies on a sequencing operator \cite{BLB19}.

Our specification of the behaviour of an RTM seems to make essential use of all the constructions of the $\pi$-calculus. Interesting future work would be to consider the various subcalculi of the $\pi$-calculus and determine to what extent these are behaviourally complete. It could then also be worthwhile to consider deterministic reactive Turing machines. Another interesting approach to simulate reactive Turing machines in the $\pi$-calculus could proceed via a universal process of the latter \cite{Fu17}.

We have also established that the $\pi$-calculus is orbit-finitely executable up to branching bisimilarity, by associating with every $\pi$-calculus process an orbit-finite RTM that simulates it. The simulation is non-deterministic and introduces divergence. We leave it as an open problem whether there exists a simulation that does not introduce divergence and proves that the $\pi$-calculus is executable up to divergence-preserving branching bisimilarity. It is established in \cite{BLvT2013} that every boundedly branching computable transition system is finitely executable up to divergence-preserving branching bisimilarity and that every effective transition system is finitely executable up to the divergence-insensitive variant of branching bisimilarity. Similar general characterisations of the notion of orbit-finite executability may be needed for solving the aforementioned open problem.

The generic specification of the behaviour of a Turing machine tape, presented in Section~\ref{subsec:tape}, does not rely on a finiteness assumption regarding the set of tape symbols; in fact, any $\pi$-calculus name can be stored on tape. The specification of the control process of an RTM in Section~\ref{subsec:fincontrol}, however, does essentially rely on the RTM having finitely many states, finitely many tape symbols and finitely many action symbols, for it has summations indexed by the sets of states, tape symbols and transitions of the RTM to be simulated. Thus, orbit-finite RTMs can be simulated by a variant of the $\pi$-calculus that allows summations indexed by orbit-finite sets. We conjecture that, on the one hand, infinite summations are essential for a simulation up to divergence-preserving branching bisimilarity, whereas, on the other hand, up to the divergence-insensitive variant of branching bisimilarity they are not.

\section*{Acknowledgments}
\noindent The first author thanks Jos Baeten for being an inspiration, for instigating the research on the integration of automata theory and concurrency theory, and for the pleasant collaboration on this and other topics over the years. Furthermore, the authors thank the anonymous reviewers for their suggestions, which led to improvements of the presentation.

\bibliographystyle{alpha}
\bibliography{main}
\end{document}